\documentclass[fleqn,usenatbib]{mnras}
\usepackage{newtxtext,newtxmath}

\usepackage[T1]{fontenc}
\usepackage{ae,aecompl}
\usepackage[dvipsnames]{xcolor} 

\usepackage{graphicx}	
\usepackage{amsmath}	
\usepackage{amssymb}	

\def\simeq{
\mathrel{\raise.3ex\hbox{$\sim$}\mkern-14mu\lower0.4ex\hbox{$-$}}
}

\def\msun{{\rm M_{\odot}}}
 
\def\le{{L_{\rm Edd}}}

\def\del#1{{}}
\def\ltsima{$\; \buildrel < \over \sim \;$}
\def\simlt{\lower.5ex\hbox{\ltsima}}
\def\gtsima{$\; \buildrel > \over \sim \;$}
\def\simgt{\lower.5ex\hbox{\gtsima}}
\def\sgra{Sgr~A$^*$}
\def\deg{^{\circ}}

\newcommand{\MTc}[1]{#1}

\title[AGN feeding by dynamical perturbations]{Feeding of active galactic nuclei by dynamical perturbations}

\author[M. Tart\.{e}nas \& K. Zubovas]{
Matas Tart\.{e}nas$^{1,2}$\thanks{E-mail: matas.tartenas@ff.stud.vu.lt},
Kastytis Zubovas$^{1,2}$
\\
  $^{1}$Center for Physical Sciences and Technology, Saul\.{e}tekio av. 3, Vilnius LT-10257, Lithuania \\
  $^{2}$Vilnius University Observatory, Saul\.{e}tekio av. 3, Vilnius LT-10257, Lithuania\\
}

\date{Accepted XXX. Received YYY; in original form ZZZ}

\pubyear{2019}

\begin{document}
\label{firstpage}
\pagerange{\pageref{firstpage}--\pageref{lastpage}}
\maketitle

\begin{abstract}

There possibly was an AGN episode in the Galactic Centre about 6 Myr ago, powerful enough to produce the {\em Fermi} bubbles. We present numerical simulations of a possible scenario giving rise to an activity episode: a collision between a central gas ring surrounding the supermassive black hole (SMBH) and an infalling molecular cloud. We investigate different initial collision angles between the cloud and the ring. We follow the hydrodynamical evolution of the system following the collision using {\sc Gadget-3} hybrid N-body/SPH code and calculate the feeding rate of the SMBH accretion disc. This rate is then used as an input for a 1D thin $\alpha$-disc model in order to calculate the AGN luminosity. By varying the disc feeding radii we determine the limiting values for possible AGN accretion disc luminosity. Small angle collisions do not result in significant mass transport to the centre of the system, while models with highest collision angles transport close to $40\%$ of the initial matter to the accretion disc. Even with ring and cloud masses equal to $10^4 \, \msun$, which is the lower limit of present-day mass of the Circumnuclear ring in the Galactic Centre, the energy released over an interval of 1.5 Myr can produce $\sim 10\%$ of that required to inflate the {\em Fermi} bubbles. If the gas ring in the Galactic Centre 6 Myr ago had a mass of at least $10^5 \, \msun$, our proposed scenario can explain the formation of the {\em Fermi} bubbles. We estimate that such high-impact collisions might occur once every $\sim 10^8$~yr in our Galaxy.

\end{abstract}




\section{Introduction} \label{intro}


It is now well established that there is a supermassive black hole (SMBH) at the centre of the Milky Way Galaxy, which coincides with the radio source \sgra \, \citep{SizeShapeSgrA}. The mass of the black hole, determined from the orbits of nearby S stars, is $4.02\pm0.16\pm0.04\times 10^6\msun$ \citep{BHMASS}. 
At the moment the SMBH is inactive but there were at least two activity periods in the comparatively recent past \citep{traces2012}. The larger of these two episodes happened $\sim 6$~Myr ago and probably involved \sgra \, reaching its Eddington luminosity $\le \approx 5.2\times10^{44}$ erg s$^{-1}$. This activity period is most likely responsible for the formation of the {\em Fermi} bubbles \citep{kzbubles2012} - the huge gamma-ray emitting structures, stretching from the centre for about 10 kpc perpendicular to the Galactic plane \citep{fermior}. \MTc{Recent observations expanded this picture, revealing smaller-scale structures, such as `X-ray chimneys' - channels from the galactic centre (GC) to the {\em Fermi} bubbles \citep{XRAYCHIMNEY}, - and two $430$-parsec-tall `radio bubbles' \citep{430RADIOBUBBLES}}. Another, less direct evidence of this activity episode are two observed rings of stars
of comparable age concentrated in the central 0.5 pc \citep{YoungDiscEvidence}. 
The presence of these rings suggests that there was a large amount of gas surrounding \sgra\, around that time. 

Observations of the interstellar medium (ISM) close to the GC reveal a ring of molecular gas surrounding the \sgra \, radio source. Known as the Circum-nuclear ring (CNR), it has an inner radius of about $2\,$pc \citep{gas}. Hydrodynamical simulations \citep{CNRform2015, CNRform2018} show that the CNR-like feature can form if there is an infall of matter to the central region, for example from a tidally disrupted molecular cloud (MC).

If clouds occasionally fall into the GC, as hinted in \cite{CNRfeedl}, we would expect collisions between infalling matter and the CNR-like ring surrounding \sgra\, to occur. Such a collision could initiate a period of nuclear activity in the Milky Way by transporting a large amount of gas toward the SMBH. Molecular clouds, gas streams and remnants of cloudlets are observed in the central few parsecs \citep{gas, cloudSurvey, cloudlets}, showing that there is an inflow of gas into the central regions \citep{CNRfeedl,CNRfeed2}. 

Infalling clouds may also result in star formation within the central parsec. Initially, the presence of young massive stars in the region was thought to be a problem \citep{gcrev}, however subsequent work showed how star formation may proceed under these conditions. The subject of star formation from a disrupted MC in a region around the SMBH was explored in \cite{StarFormationGC1}. They found that a MC falling into the SMBH would produce an eccentric and clumpy gas disc bound to the SMBH. This disc could be a site of star formation with a top-heavy initial mass function, as the small-scale clumps were not disrupted by the SMBH. A similar idea was further explored in \cite{StarFormationGC2} where it was found that the observed population of young massive stars could have originated from a collision of two MCs at 1 pc from \sgra. The initial collision resulted in the formation of rings which over time became self-gravitating and warped, producing stellar populations of differing parameters. \MTc{However, this study is based on somewhat contrived initial conditions - the presence of two clouds on almost identically opposite orbits. A direct collision of a MC and the central SMBH would result, depending on the impact parameter, in a period of increased accretion and build-up of an accretion disc which could possibly form stars \citep{Alig1}. Exploring a similar setup, \cite{LUCASMCINFALL} found that stars formed at sub-parsec distances from the SMBH. \cite{Alig2} modelled a collision with a pre-existing gas ring and found that it is capable of producing two inclined accretion discs.}

In this paper, we consider the collision between a gas cloud and a CNR-like gas ring and its effects on nuclear activity. We model the scenario using a set of simulations run with the hybrid N-body/SPH code Gadget-3 \citep[an updated version of the publicly-available Gadget-2,][]{GADGET2005}. The results are then used as input in a separate accretion disc model.
This approach allows us to produce a more realistic light-curve of the AGN episode than only using the accretion rate on to the SMBH in the hydrodynamic simulation, which in turn improves the calculation of the total released energy and the estimate of the episode duration. With a suite of models we test the dependence of main results on the initial collision angle between the cloud and the ring, the stochastic variation of particle positions, and the location of gas infall on to the accretion disc. 
We find that infall at very large angles can produce enough accretion to power the formation of the {\em Fermi} bubbles. The exact values of the angle depend on the mass of the ring, but are $\gamma > 150 \deg$ even for the highest physically plausible mass $M_{\rm ring} = 10^6 \, \msun$.  A collision of this scale is estimated to occur once per $60 - 140$~Myr. \MTc{The total accreted mass has little dependence on the cloud-to-ring mass ratio or the radius of the cloud.} We analyze the morphology of resulting structures and find that the mass and size of the present-day CNR can inform us about the likely properties of a past collision event. We also discuss the applicability of our results to other galaxies, where similar CNR-like gas rings are commonly observed. 

\begin{figure*}
    \begin{centering}
        \centering
        \begin{centering}
        \includegraphics[width=0.8\textwidth]{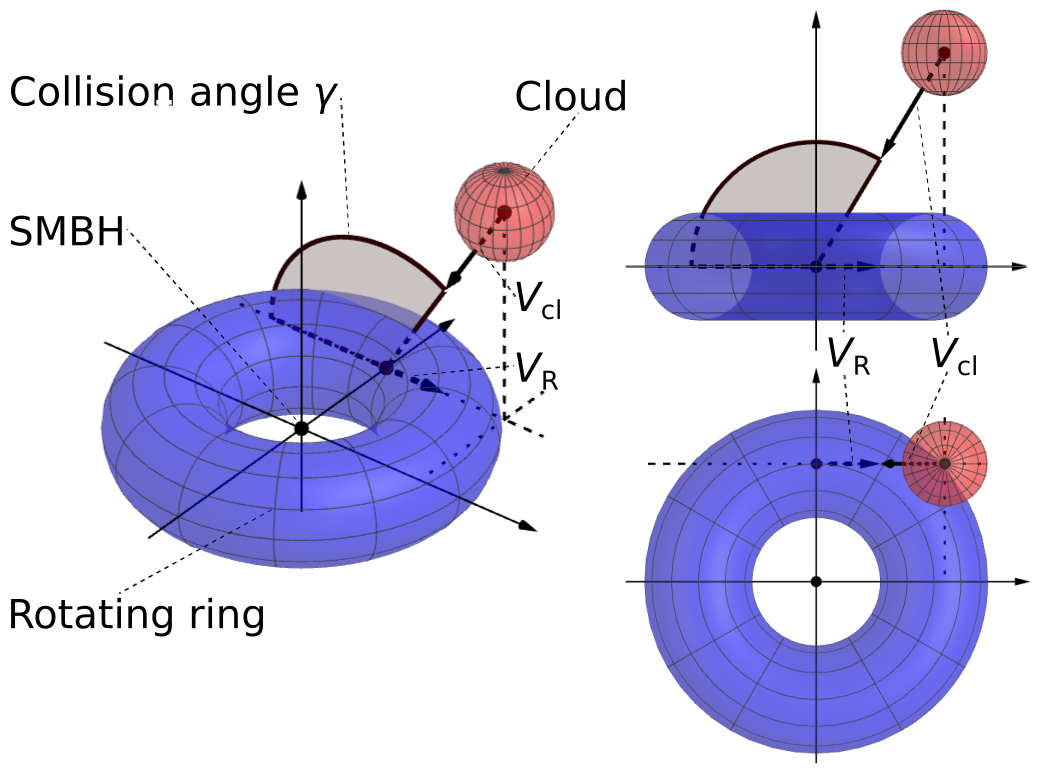}
        \end{centering}
        \caption{\label{drawing} A schematic drawing of the initial system. The CNR-like gas ring is represented by the blue torus and the infalling clous - as a red sphere. Vectors represent initial velocities: $V_{\rm cl}\,$ for cloud and $V_{\rm R}\,$ for the tangential component of the rings rotation.}
        \end{centering}
\end{figure*}

A more detailed description of the models used is given in section \ref{PaN}. Model results are presented in section \ref{results}, with sections \ref{morp}-\ref{GasT} focusing on the SPH simulation results, while sections \ref{ACC}-\ref{totalEnergy} the focus is on the results of accretion modelling. In section \ref{Discussion} we discuss the implications of our results and some of the more problematic parts of the model, followed by a conclusion in section \ref{Conclusions}.

\section{Physical and numerical model} \label{PaN}

Our simulation is composed of two distinct parts. First we use Gadget-3 \citep{GADGET2005} to model the collision between the CNR and a gas cloud. This simulation allows us to investigate the morphology and kinematics of the resultant system. We also calculate the feeding rate for the SMBH accretion disc. We then use this feeding rate as an input to a 1D $\alpha$ accretion disc model to obtain a more realistic picture of the probable activity episode duration and luminosity evolution over time.

\subsection{Hydrodynamic model} \label{hydroMod}

We simulate a collision between the ring and the cloud with the three-dimensional hybrid SPH/N-body code Gadget-3, an updated version of the code presented in \cite{GADGET2005}. We use an SPH implementation with a higher order dissipation switch \citep[SPHS;][]{SPHS2012}. For SPHS, the appropriate smoothing kernel is a Wendland function C$^2$ \citep{kernel} with neighbour number $N_{\rm neigh} = 100$. Each of our models has $N_{\rm part} = 5\times10^5$ particles with mass $m_{\rm SPH} = 0.04 \, \msun$. The resolved mass is $M_{\rm res} = N_{\rm neigh} m_{\rm SPH} = 4 \textrm{M}_{\odot}$. 

The system initially consists of three components. At the origin, there is the central black hole with mass $M_{\rm bh} = 4 \times 10^6\, \textrm{M}_{\odot}$. Gas particles that come closer than  $R_{\rm bh} = 0.01$~pc to the black hole particle and are gravitationally bound to it are swallowed and form the central accretion disc. The SMBH is surrounded by a toroidal gas ring with $M_{\rm ring} = 10^4 \textrm{M}_{\odot}$, $r_{\rm in} = 1.5$ pc and $r_{\rm out} = 4$ pc. These properties are similar to those of the CNR \citep{gas}; however, the chosen mass of the ring is a lower limit, and various observational estimates can be up to two orders of magnitude higher \citep{maxmass}. Finally, a molecular cloud with $M_{cl} = M_{ring} = 10^4 \textrm{M}_{\odot}$, $r_{\rm cl} = 1$~pc is placed 6  pc away from the origin. Clouds of similar size and mass have been observed in the GC \citep{cloudSurvey}.

We use potential given by the sum of the gravitational potential of the central SMBH and the isothermal potential:
\begin{equation}
   \phi = -\frac{\textrm{G}M_{\rm{bh}}}{r} - 2\sigma^2 \log \frac{r}{r_0},
\label{pot}
\end{equation}
where the first term is the gravitational potential of a point mass and the second is an isothermal potential, with velocity dispersion $\sigma = 100$~km$\,$~s$^{-1}$; r$_{0}$ is an arbitrary large constant. \MTc{Our chosen potential corresponds to an enclosed mass $M_{\rm enc} = M_{\rm bh}$ at $R_{\rm enc} = 0.8$~pc. This is somewhat smaller than the $r_{\rm NSC} = 3.5$~pc radius of the Nuclear star cluster, which has a mass similar to $M_{\rm bh}$ \citep{NuclearStarPot}. However, our potential corresponds to stars, stellar remnants and dark matter, therefore must have a higher enclosed mass than just that of the stars.}

Ring particles move in circular orbits with speeds $v_{R1.5} \sim 181$ km s$^{-1}$ at the inner edge and $v_{R4} \sim 160$ km s$^{-1}$ at the outer. The cloud is set on a collision course on a parabolic orbit set to pass through the middle of the ring at a point $2.75$~pc away from the origin along the X axis. The initial velocity of the centre of the cloud is $v_{\rm cl} = 220$km s$^{-1}$. 

In addition to orbital velocities, all particles are given velocities from a turbulent velocity field. This is created based on the example of \cite{Turbulencija}, with velocity amplitude $\sigma_{\rm turb} \sim 37.5$ km s$^{-1}$. The Kolmogorov power spectrum of the velocity field for homogeneous and incompressible turbulence is 
\begin{equation}
    P_\nu  \equiv  \langle |v_k|^2 \rangle \propto k^{-11/3},
\end{equation}
where k is the wave number. The flow is divergence-free (i.e. turbulence is purely solenoidal) in incompressible fluid and so we can define a vector potential, \textbf{A}. The components of \textbf{A} are described by a Gaussian random field, and velocity field can be derived by \textbf{v} = $\nabla$ $\times$ \textbf{A}. By dimensional arguments, the new power spectrum is
\begin{equation}
   \langle |A_k|^2 \rangle \propto k^{-17/3}.
\end{equation}
The dispersion of |\textbf{A}|  for a field point diverges, so a cut-off wave number $k_{\rm min}$ is introduced to ensure convergence and power spectrum is redefined as
\begin{equation}
    \langle |a_k|^2 \rangle = C(k^2 + k_{\rm min}^2)^{-17/6},
\end{equation}
Where C is a normalisation constant. Since $k \propto 1/L$, the physical interpretation of $k_{\rm min}$ is that $R_{\rm max} \simeq k_{\rm min}^{-1}$ is the largest scale on which the turbulence is driven. The field is generated by first sampling the vector potential \textbf{A} in Fourier space, drawing amplitudes of each component at points ($k_x, k_y,k_z$) from a Rayleigh distribution with variance given by $\langle | A_k|^2 \rangle $ and assigning uniformly distributed phase angles between 0 and 2$\pi$. Then the  curl is taken
\begin{equation}
    \boldsymbol{v}_k = i  \boldsymbol{k} \times \boldsymbol{A}_k
	\label{eq:pspec}
\end{equation}
to obtain the velocity field  components in Fourier space. Fourier transform is then taken to get the velocity field in real space. We use a grid of $64^3$ cells when generating the statistical realization of the field. Finally, individual particle velocities are interpolated from nearby grid cell values.

In order to speed up the simulations and to make the results independent of actual masses of the ring and cloud, but only on their mass ratio, we use a $\beta$-prescription \citep{BETA2011} for gas cooling and turn off gas self gravity. Observations of the CNR show that its gas is probably not self-gravitating. Its probable mass is an order of magnitude smaller than the self-gravity threshold $M_{\rm virial} \sim 5.4\times10^6\,\msun$ \citep{CNRVirialmass}, therefore we turn off gravity between gas particles in our simulation, leaving only the background and SMBH gravitational potentials.
We parameterise cooling with the $\beta$-prescription, where $\beta$ is a constant coefficient that ties the cooling time-scale with dynamical time-scale: $t_{\text{cool}} = \beta_{\text{cool}} t_{\rm d}$, where the dynamical time is given by $t_{\rm d} = r/\sqrt{2}\sigma$. In our model $\beta = 0.1$, i.e. cooling is rather efficient. We further explore the importance of cooling time in section \ref{Betarez} where the results of simulations with varied $\beta$ are presented.

In total, we simulate 52 collision scenarios with different cloud collision angles $0^\circ < \gamma < 180^\circ$, incrementing them by 15$^{\circ}$ steps. This angle is the angle between the initial orbital angular momentum vectors of the cloud and the ring. Setting $\gamma$ to 0 results in a prograde collision and $\gamma = 180^{\circ}$ produces an ideally retrograde collision. For each value of $\gamma$ we run four simulations with stochastically different initial gas particle distributions in order to analyze variations due to possible chaotic aspects of the system's evolution.

%
\begin{figure*}
    \begin{centering}
    \includegraphics[width=\textwidth]{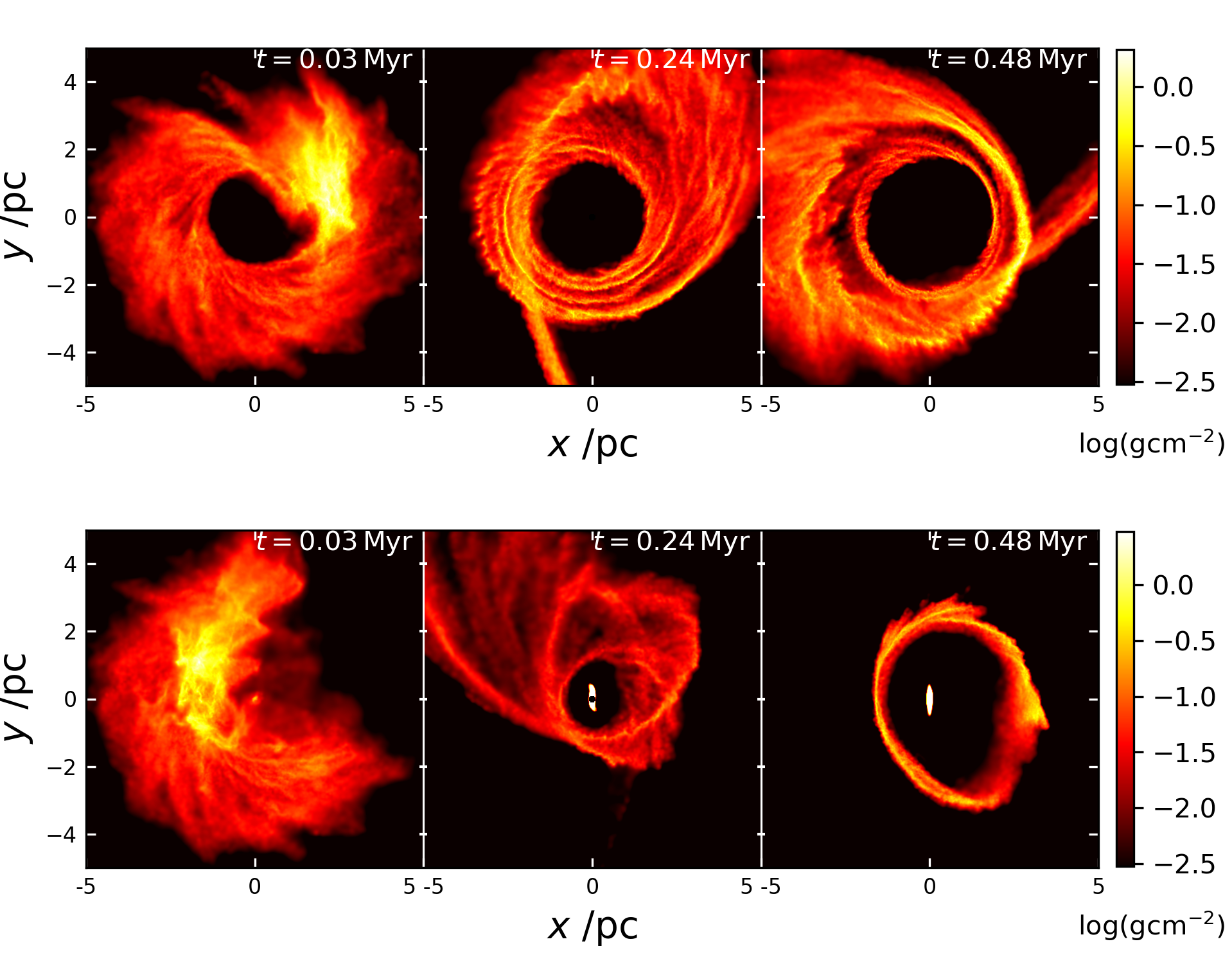}
		\caption{\label{fig:D_map_evo}Evolution of models with collision angle $\gamma$ set to $15^{\circ}$ (top) and $165^{\circ}$ (bottom). Moment just after the initial contact is shown on the left ($t = 0.03$). 
		}    
    \end{centering}   
\end{figure*}

\subsection{Accretion disc model} \label{sec:accdiscmodel}

The SMBH accretion disc is too small to be resolved in our hydrodynamical simulation. However, matter transport through the disc by viscous torques is important and could extend the duration of the activity period compared with the period over which material is fed to the SMBH particle in the hydrodynamic simulation. In order to more accurately portray the possible luminosity evolution of the system we use an idealised 1D thin $\alpha$-prescription accretion disc model based on \cite{Pringle1981}. 

The model is an Eulerian integration scheme for the thin-disc equations such as described in \cite{ShakuraSunyaev} and, e.g., Chapter 5 of \cite{frank_king_raine_2002}. The disc consists of a number of annuli rotating in Keplerian orbits around a central mass. Material is transported through the disc toward the centre by the action of viscous torques, emitting thermal radiation. The mechanism of radiation emission is usually parameterised as being generated by the dissipation of viscous stress. The shear viscosity $\nu$ is parametrized by an $\alpha$ prescription \citep[cf.][]{ShakuraSunyaev}: $\nu = \alpha c_{\textrm{s}} H$, where $c_{\rm s}$ is the sound speed and $H$ is the vertical scale height of the disc. We assume that each annulus is a separate black body for the purpose of luminosity calculation, so the full spectrum is given by the superposition of the spectra from each annulus and the luminosity at any given time is:
\begin{equation}
    L = \sum_i L_i = \sigma_{\rm SB} \sum_i A_i T_i^4,
    \label{Lint}
\end{equation}
where $\sigma_{\rm SB}$ is the Stefan-Boltzmann constant and $A_i$ and $T_i$ are the area and temperature of the $i$th annulus, respectively.

Our computational grid stretches from the ISCO ($R_{\rm ISCO} = 3\,R_{\rm{s}}$) to the outer edge at $R_{\rm out} = 26107\,R_{\rm{s}} = R_{\rm{bh}} = 0.01$~pc. It consists of 151 annuli with logarithmically increasing width: $l_{i+1}/l_i \simeq 1.06$. The width of the innermost annulus is $l_0 \approx 0.18\, R_{\rm s}$ and the outermost has $l_{150} \approx 1520\, R_{\rm  s}$. 

Initially, the grid is empty. For the first $t_{\rm f} = 0.5$~Myr, the disc is fed by material that falls in from the outside; this feeding rate is taken from the hydrodynamic simulation by measuring the matter infall rate through the sink radius of the SMBH particle. Gas surface density in the disc is evolved by numerical integration of the evolution equation:
\begin{equation}
    \frac{\partial \Sigma}{\partial t} = \frac{3}{R}\frac{\partial}{\partial R} \left\{ R^{1/2}\frac{\partial}{\partial R} \left[\nu\Sigma R^{1/2}\right] \right\},  
    \label{dSigmadt}
\end{equation}
where $\Sigma$ is the surface density and $R$ is the radial coordinate. 

An uncertainty arising from our two-model approach is that there is no clear way to know where in the disc the mass should be injected. It arises because the smoothing lengths of the SPH particles falling on to the SMBH are typically larger than $R_{\rm bh} = 0.01$~pc. We try to circumvent this problem by varying the feeding radius $R_{\rm f}$. We choose five feeding radii: $R_{\rm f}/R_{\rm s} = 300, 5725, 11150, 16575$ and $22000$. In reality, there would be a spread of radii at which material falls on to the disc, but we expect that our method provides reasonable upper and lower limits to SMBH accretion rates. Another drawback is the lack of feedback: since we run the 1D code independently of the hydrodynamical one, the latter cannot affect the former in any way, but this is not critical, since the momentum of the AGN wind at its peak is still much smaller than the weight of the central disc (see Section \ref{feedback}).

\section{Results }\label{results}

\subsection{Morphology of the resultant system}\label{morp}

Given our setup, the cloud reaches the CNR-like ring in $\sim 20$~kyr. The initial system is perturbed by the collision; its subsequent evolution depends strongly on the initial collision angle.

Fig. \ref{fig:D_map_evo} shows three snapshots of evolution of two simulations as seen by an observer located on the Z axis. After the initial collision, the system with the collision angle $\gamma = 15^{\circ}$ (top panels) is only slightly perturbed in contrast with the system with $\gamma = 165^{\circ}$ (bottom panels) where the impact disrupts the initial system completely. In the first case, almost all the gas settles onto the perturbed initial ring, since the gas cloud does not oppose the rotation of the ring and even accelerates some of the ring particles. In the second case, the cloud opposes the rotation of the ring and the collision slows down some of the gas. Because of this, a large fraction of the initial gas mass is transported to the centre of the system, where it forms a central disc perpendicular to the rotation plane of the initial ring and the remaining matter settles into a somewhat smaller and significantly narrower ring. In both cases, the rings are patchy, made of narrow filaments.

For further analysis, we define two types of resultant structures: discs and rings. A disc is a structure that extends outward from the accretion radius $R_{\rm bh}$ and has a clear outer edge. A ring is a structure bounded by sudden density drops at both the inner and outer edges. We use radial gas density profiles and density maps to label individual gas particles as belonging to either a disc, a ring, or neither. 
In some cases, these distinctions are somewhat subjective, but they help us investigate the evolution of the system. Furthermore, a disc and a ring should be clearly distinguishable with spatially-resolved observations, therefore understanding the structures that form under different circumstances allows us to predict the observational signatures of the modelled events. In many cases there are also streams of matter that did not yet settle on to a ring given our simulation time. We did not consider matter in the stream as part of the ring, but it could be reasonably assumed that most of the matter would eventually join the reformed ring leaving only a few straggler particles too far from the initial system to be considered.

\begin{figure}
	\includegraphics[width=\columnwidth]{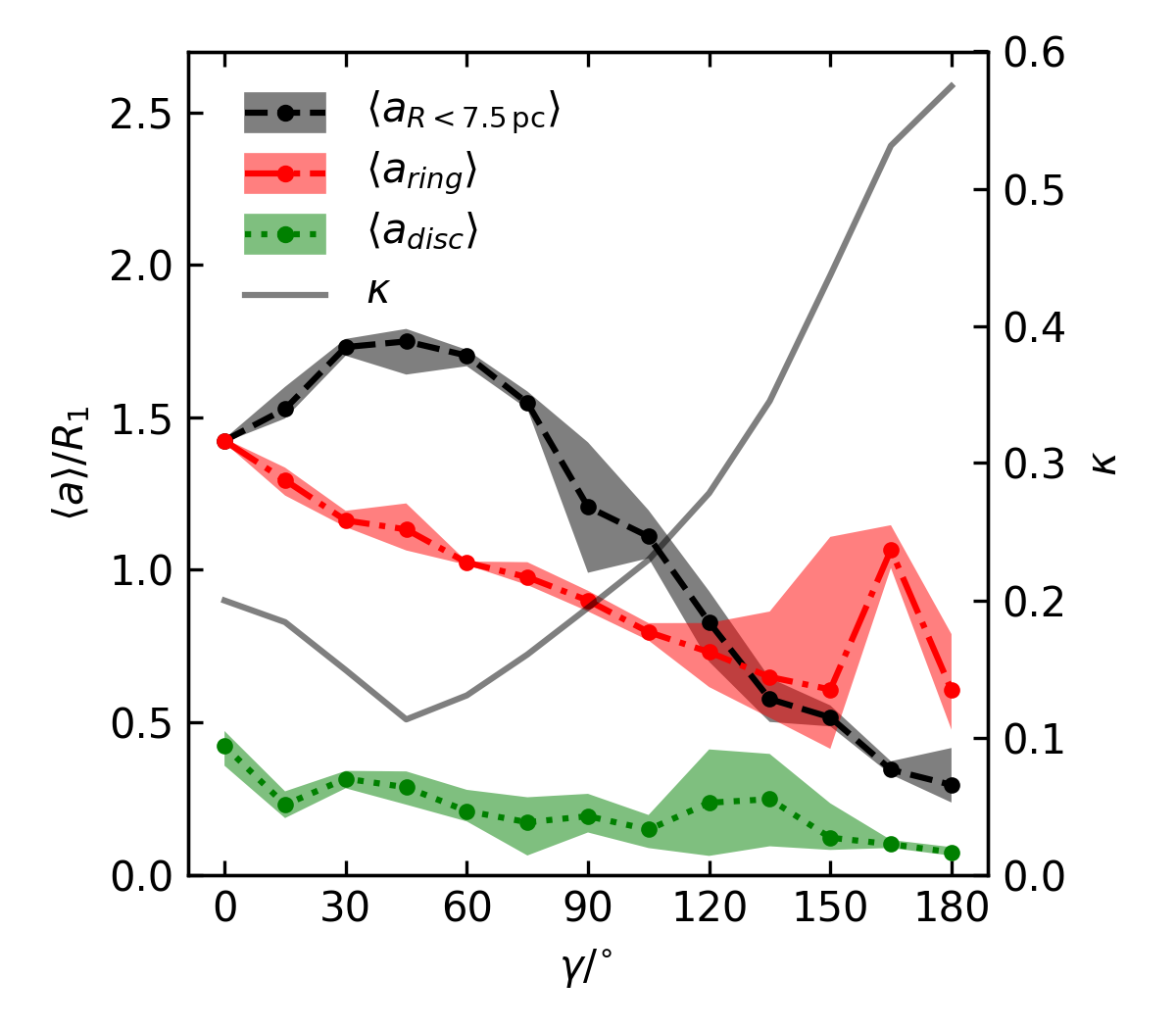}
	\vspace{-0.75cm}
    \caption{The average circularization radii of gas at $t = 0.5$~Myr in simulations with different initial collision angles. Black line: all gas within $7.5$~pc; red line - the resultant rings, green line - the central disc. Shaded regions show the variance among stochastically different models with the same collision angle. The grey line (scale on the right) shows the fraction $\kappa$ of the ring contacted by the cloud during initial passage (eq. \ref{n_mass}).}
    \label{fig:circOrb}
\end{figure}

\begin{figure}
	\includegraphics[width=\columnwidth]{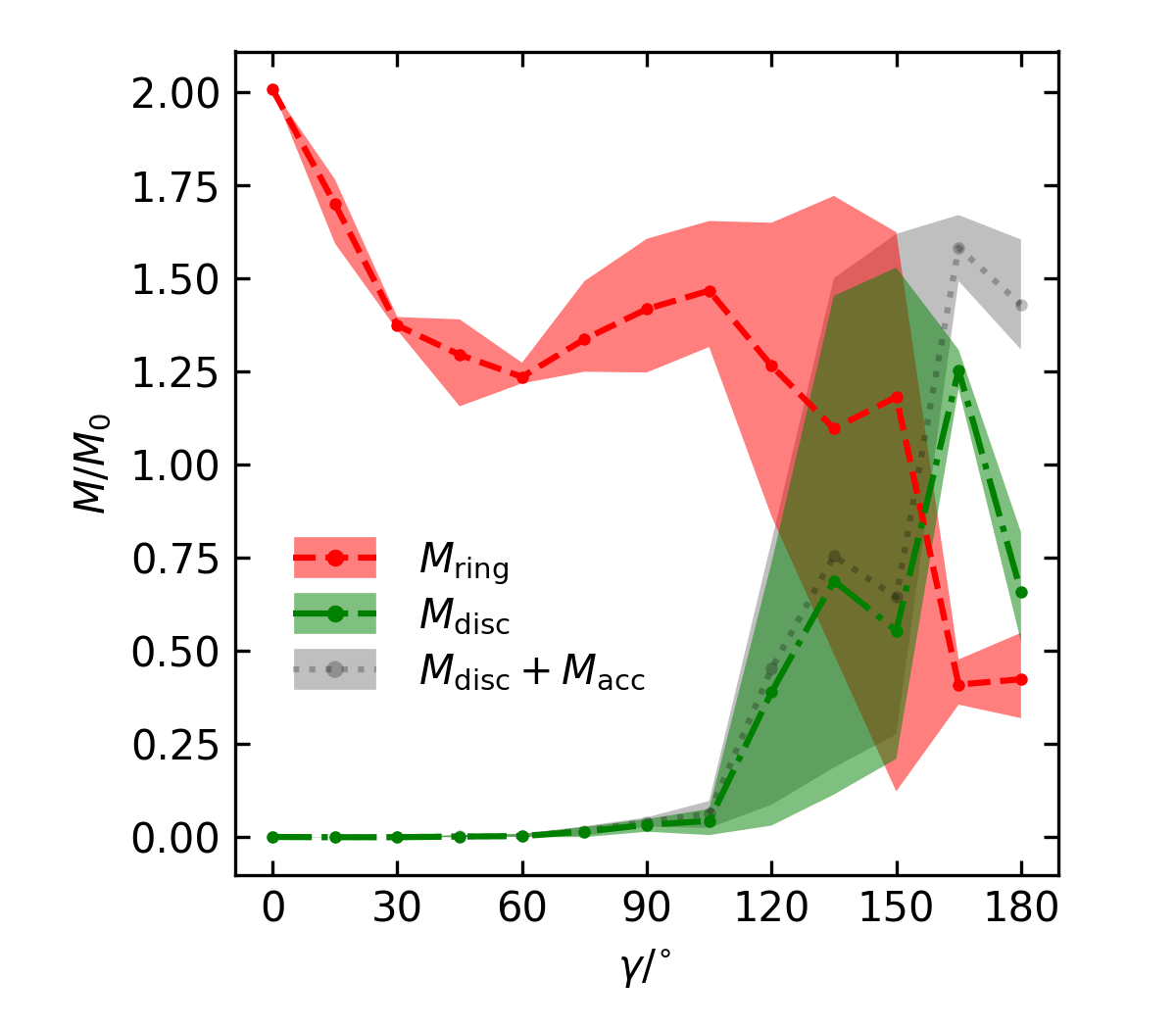}
	\vspace{-0.75cm}
    \caption{\label{fig:structMass}Average masses of the resultant rings (red) and central discs (green) in models with different initial collision angle $\gamma$ expressed in ratio with the initial ring mass $M_{0} = 10^4\textrm{M}_{\odot}$. Grey line shows the sum of disc mass and mass accreted by the SMBH particle. Shaded areas show variations between models with the same $\gamma$. 
    }
\end{figure}

\begin{figure*}  
    \begin{centering}
        \includegraphics[width=\textwidth]{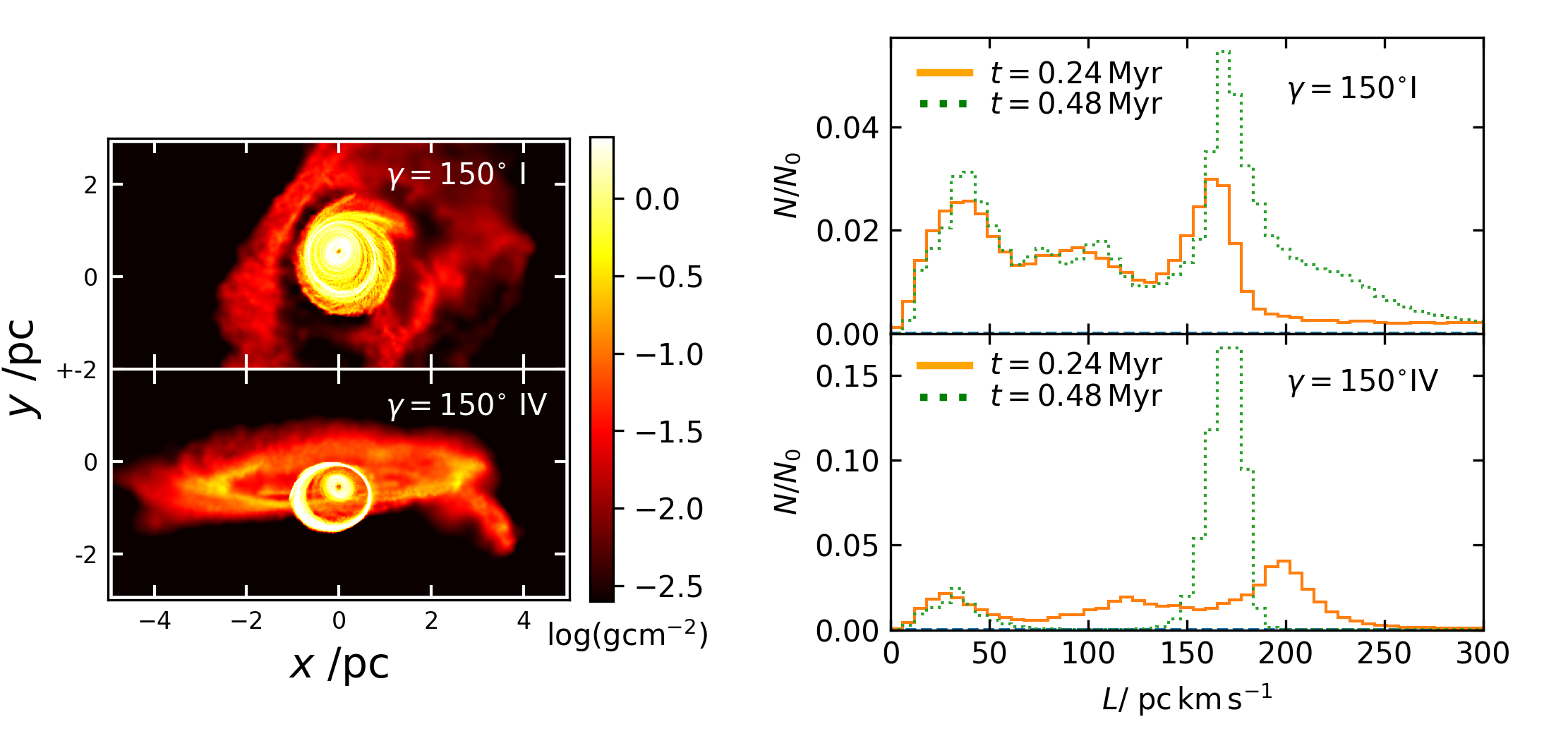}
		\caption{ \label{fig:Dens_hist}Density maps of two simulations with $\gamma = 150 \deg$ at $t = 0.48\,$Myr (left) and  distribution of gas angular momentum per unit mass of the central region in the same simulations at times $t = 0.24\,$Myr and $t = 0.48\,$Myr (right). A peak forms in both I and IV simulations, but the one in I did not break from the central disc. 
		}    
    \end{centering}   
\end{figure*}

Broadly speaking, large-angle collisions lead to formation of more compact systems; this is clearly seen in Fig.~\ref{fig:circOrb}. Here we show the circularization radii of the defined structures (discs in green, rings in red) and all particles (in black). Lines show median values of the four stochastically different simulations at each angle and the shaded region encompasses the stochastic variations. The average circularization radii of the gas particles reaches a peak at $\gamma = 45\deg$ and is minimal at $\gamma = 180\deg$. This behaviour is a consequence of different `encounter lengths', which are the fractions of total ring and cloud material that interact directly during the first passage of the cloud through the ring. This fraction can be roughly estimated analytically:
\begin{equation}
\kappa \approx \frac{v_{\rm{enc}} t_{\rm{enc}} + 2r_{\rm{cloud}}}{C},
\label{n_mass}
\end{equation}

where $v_{\rm enc} \equiv v_{\rm{cloud}} \cos \gamma - v_{\rm{orb}}$ is the relative velocity of the cloud and ring material, $t_{\rm{enc}} \equiv l_{\gamma}/v_{\rm cloud}$ is the time required for the cloud to pass through the ring, $l_{\gamma}$ is the length of the path of the cloud's orbit inside a spiric section of the toroidal ring for a given angle $\gamma$, and $C$ is the circumference at the midpoint of the torus. $\kappa$ then shows the fraction of the ring circumference that the cloud travels while it stays in contact with the initial ring. The dependence of $\kappa$ on the collision angle is plotted as a grey line in Figure \ref{fig:circOrb}. Even though this is a very approximate estimate, its qualitative anti-correlation with the mean circularization radius is clearly evident. The peak circularization radius occurs in the simulation in which the secondary collisions do not transport a large amount of gas to the central part of the system. The stochastic variation of circularization radius is small for angles $30\deg \leq \gamma \leq 60\deg$, so this result is robust. The peak is only seen when we look at all the particles, including the ones that did not yet settle on to the central disc or rings. This may be interpreted as a consequence of the cloud carrying away some fraction of the ring material, which then stretches into elongated spiral orbits and has not settled into a ring-like structure by $t = 0.5\,$Myr. The large variance of the disc and ring radii in simulations with $120\deg \leq \gamma \leq 150\deg$ is caused by the disc and ring having a very narrow gap in these simulations, so their relative sizes vary significantly due to stochastic differences; the variance disappears when we consider all the particles.

The masses of the resultant structures are shown in Fig. \ref{fig:structMass}. In simulations where $\gamma>120^{\circ}$ almost all of the gas settles into either a ring (red line and shaded region), a disc (green line and shaded region) or feeds the central BH (accounted for as grey line and shaded region).
Most of the gas settles into rings for collision angles $\gamma <120^{\circ}$, although central discs appear in simulations with collision angles $\gamma > 60^{\circ}$. Increasing collision angles further results in systems where the distinction between the central disc and rings is less pronounced as can be noted seeing the large variance in models with $120^{\circ}<\gamma<165^{\circ}$. In the case of the most extreme collisions, most of the gas accumulates in the central disc as well as feeding the central BH. 

Some large-angle simulations produce an additional narrow ring close to and in the same plane as the central disc. These narrow rings are situated closer to the centre and are less extended than the initial ring or patchy rings that form after the collision. An example is shown in fig. \ref{fig:Dens_hist} (left panels) with their angular momentum distributions in the right panels. A ring is clearly seen in simulation IV, but not in simulation I. These rings form because gas with different angular momentum mixes radially, and the angular momentum distribution narrows due to shocks eventually producing the highly peaked distribution \citep{ACC_RING_angularmomentum}.The angular momentum distribution for  simulation IV shows that at $t\sim0.24\,$Myr the angular momentum is somewhat equally distributed but over time the distribution peaks leaving an almost empty patch between $60 \,$pc$\,$km$\,$s$^{-1}$ and $125\,$pc$\,$km$\,$s$^{-1}$. Conversely, in simulation I the angular momentum distribution peak is much less pronounced. The same process produces the narrow filaments in the extended rings, seen in fig. \ref{fig:D_map_evo}. 

\begin{figure}
	\includegraphics[width=\columnwidth]{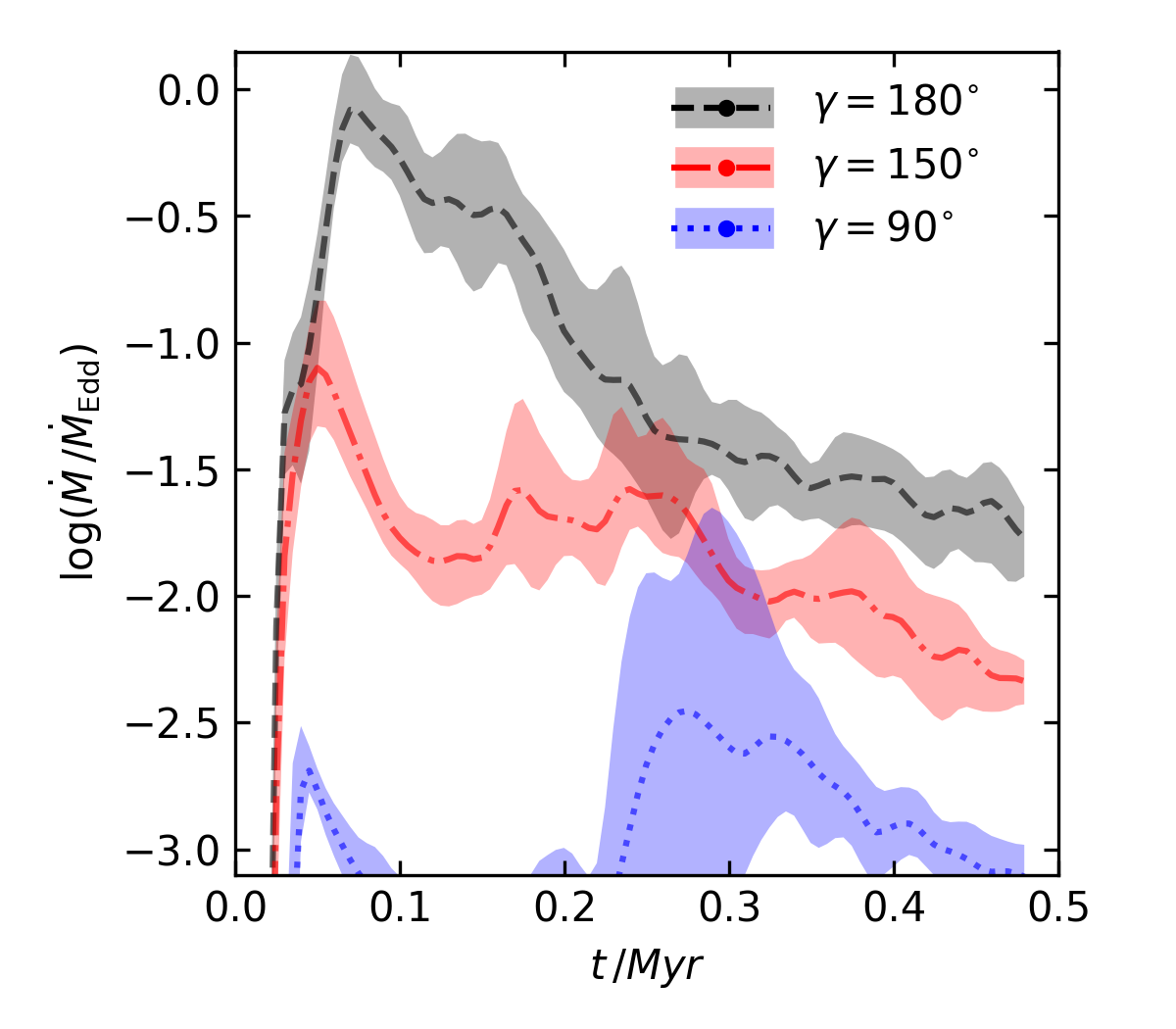}
	\vspace{-0.75cm}
    \caption{Time evolution of the average accretion disc feeding rate, scaled to the Eddington accretion rate for \sgra. Shaded regions show variations among stochastically different models with the same collision angles. The curves are smoothed by applying a weighting kernel to reduce spurious noise (see text). 
    }
    \label{fig:massTransEvo}
\end{figure}

\begin{figure*}  
    \begin{centering}
    \includegraphics[width=\textwidth]{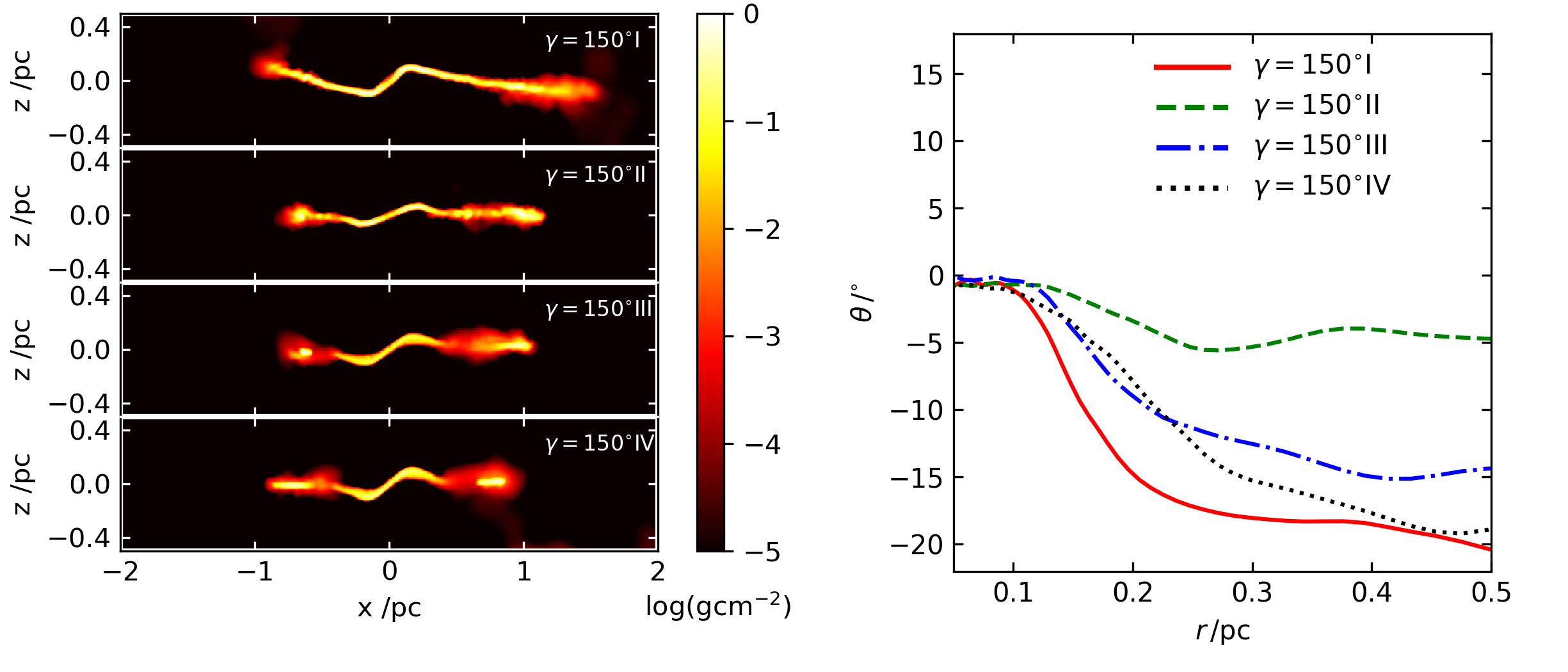} 
		\caption{ \label{fig:Slice_warp}Density wedge slices of the simulations with $\gamma = 150^{\circ}$ (left) and the tilting of the central disc for simulations with $\gamma = 150^{\circ}$ (right). The tilt does not explain the breaking of the disc as simulation I, has the largest and steepest tilt, but this did not result in a separate ring breaking off.
		}    
    \end{centering}   
\end{figure*}

The central discs in our simulation are warped (clearly seen in the left panels of fig \ref{fig:Slice_warp}). A warped disc with steep enough tilt is likely to break up \citep{Warped_disc}. However, in our simulations only a fraction of central discs break up and this does not seem to be related to the tilt: Fig. \ref{fig:Slice_warp} (right) shows the tilt angle for the four simulations; the central discs with the largest and the smallest tilts do not break up (I and II), while discs with intermediate tilts do (III, IV). So, the appearance of these dense structures seems stochastic in nature. In any case, the fact the the central disc is warped could be important for observation and the significance of narrow dense discs are explored further in the discussion.

\subsection{Gas transport}\label{GasT}

Models with larger collision angles have more gas falling into orbits closer to the black hole, which also leads to more particles passing the $r_{\rm BH} = 0.01$~pc boundary. The mass transported per unit time, scaled to the Eddington accretion rate $\dot{M}_{\rm Edd} = 9.5\times 10^{-2} \,\msun$~yr$^{-1}$, is shown in fig. \ref{fig:massTransEvo}, where we present only three representative simulations with significant gas transfer rates\footnote{In order to reduce spurious noise, the curves are smoothed using a weighting kernel $w = \lbrace 1/16,\,4/16,\,6/16,\,4/16,\,1/16\rbrace$.}.

We find that the time evolution of gas transfer rate has three qualitatively different scenarios, depending on the collision angle. With low enough angles ($\gamma < 60^{\circ}$) the initial perturbation is not strong enough to send significant amounts of gas to the centre of the system. The feeding rate is negligible in this case. Collisions with intermediate angles $60^{\circ} < \gamma < 120^{\circ}$ are able to send some of the gas to the centre of the system increasing the feeding rate at the beginning, but the collision is not strong enough to completly destroy the initial ring. The elongated remnants of the cloud and the initial ring fall into the perturbed initial ring and the newly formed central disc producing a secondary peak in accretion seen after $\sim 0.22\,$Myr, which is $\sim 20$ dynamical times of the original ring after the initial collision. 
Larger-angle collisions ($\gamma > 120^{\circ}$) result in a steep increase in feeding rate as the initial system is mostly destroyed, sending large amounts of gas to the centre passing the Eddington limit in extreme cases. The secondary peak at $t \sim 0.22$~Myr becomes less pronounced as the angle increases, since more extreme collisions result in much less material escaping as a stream that can collide with the ring during fallback.

\subsection{Accretion modelling }\label{ACC}

Using the 1D accretion disc model (see Section \ref{sec:accdiscmodel}), we investigate the luminosity evolution of the system. Only the results of SPH simulations with the three largest resulting feeding rates were used as input: those with $\gamma = 150, 165$ and $180$ degrees. Each input was used in five accretion disc simulations with different feeding radii, giving a total of 15 simulations. 

\begin{figure}
	\includegraphics[width=\columnwidth]{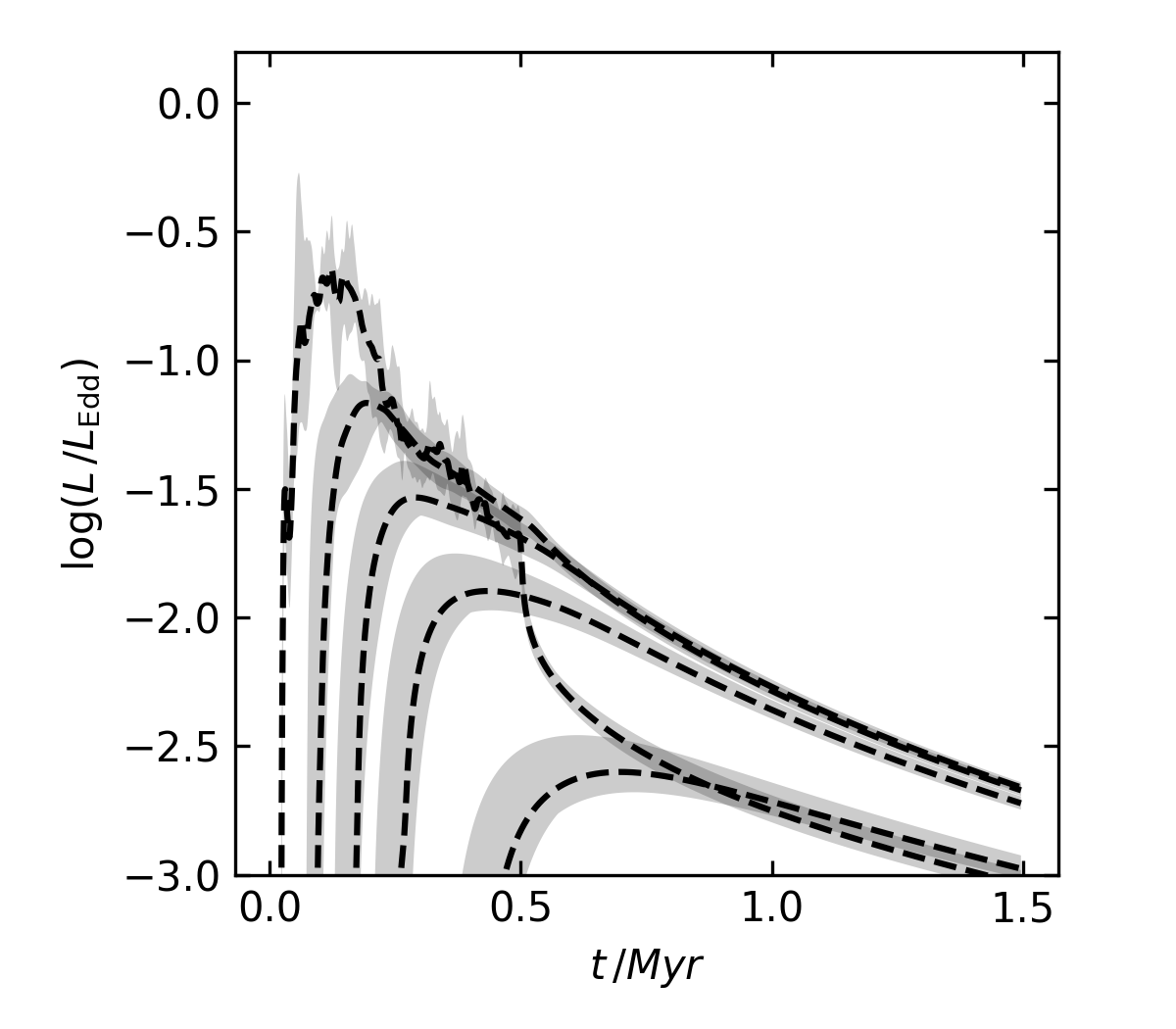}
	\vspace{-0.75cm}
    \caption{Luminosity evolution of the accretion disc with $\gamma = 180^{\circ}$. From left to right, the lines represent models with different disc feeding radii $R_{\rm f}/R_{\rm s} = 300, 5725, 11150, 16575, 22000$.}
    \label{fig:lumn}
\end{figure}

Calculated values of the luminosity evolution with feeding rates from models with $\gamma = 180^{\circ}$ are shown in fig. \ref{fig:lumn}. The simulations differ in the disc feeding radius, with $R_{\rm f}/R_{\rm s} = 300, 5725, 11150, 16575, 22000$ from left to right. Each simulation was run for 1.5 Myr to include the extended period of accretion as the disc is cleared out by viscous torques.

As expected, the peak luminosity is lower, and is reached later, the further from the centre the accretion disc is fed, mainly because disc viscosity has more time to spread the incoming material into a wider disc. In fact, the accretion rate in simulations with $R_{\rm f} = 300$ follows the evolution of the feeding rate quite closely, with an additional rapidly-dropping tail after feeding is turned off. It is worth noting that if we take the usual AGN cutoff luminosity at $L > 0.01 L_{\rm Edd}$, the model with the smallest $R_{\rm f}$ gives a shorter AGN episode duration than models with $R_{\rm f}/R_{\rm s} = 5725, 11150, 16575$, since the accretion rate gets much more smoothed out in those cases.Finally, when $R_{\rm f} = 22000$, much of the matter escapes the accretion disc without producing significant luminosity.

\subsection{Total energy release} \label{totalEnergy}

\begin{figure}
	\includegraphics[width=\columnwidth]{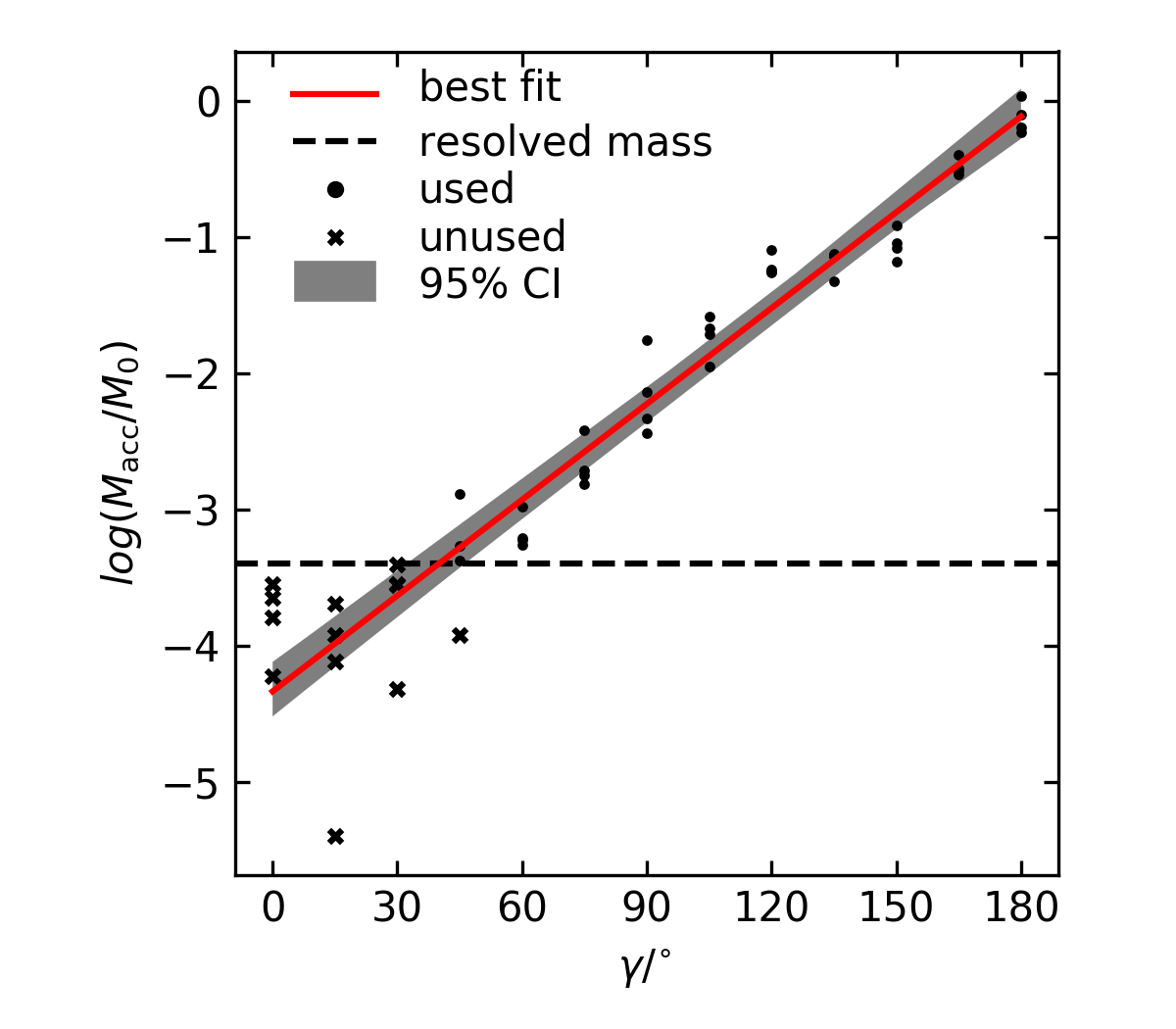}
    \vspace{-0.75cm}
    \caption{ The total mass transported to the central accretion disc over about 0.5 Myr. Red line is the best fit to the data of individual simulations (points; crosses indicate simulations that were not used for fitting, since the total transferred mass is smaller than the resolved mass limit), and the grey shaded area is the 95\% confidence limit on the line parameters.}
    \label{fig:totalMassTrans}
\end{figure}

Fig. \ref{fig:totalMassTrans} shows the total mass that feeds the BH (passes the 0.01 pc boundary) over the 0.5~Myr simulation time. Each point represents a single simulation. The total accreted mass closely follows an exponential dependence on collision angle:
\begin{equation}
    \log{M_{\rm acc}/M_{0}} \approx 2.34^{+0.14}_{-0.15} \times10 ^{-2} \gamma - 4.34^{+0.22}_{-0.19}
	\label{eq:fitted}
\end{equation}
When determining the parameters\footnote{The parameters were determined by fitting a line on a dataset generated by bootstrapping the simulation data} (slope and intercept) of the best-fit relation, points where the total mass transfer to the SMBH particle was $M_{\rm tot} < 100 m_{\rm SPH}$, which is the mass resolution of our hydrodynamical simulation, were not considered; the cutoff is marked with a horizontal dashed line. The best fit line is shown in Fig. \ref{fig:totalMassTrans} in red, with the shaded grey area encompassing a $95\%$ confidence interval of possible slopes. 

The fitted relation (eq. \ref{eq:fitted}) allows us to estimate the possible total energy output $E_{\rm{tot}}$ of the whole activity period given the angle of the collision and the initial mass of the system using:
\begin{equation}
E_{\rm tot} = \eta c^2 M_{acc} = 1.8 \times 10^{57} \eta_{0.1} M_4 {\rm erg}, 
\label{Energy_eq} 
\end{equation}
where $\eta \equiv 0.1 \eta_{0.1}$ is the radiative efficiency and $M_{\rm acc} \equiv 10^4 M_4 \, \msun$ is the total accreted mass.

\begin{figure}
	\includegraphics[width=\columnwidth]{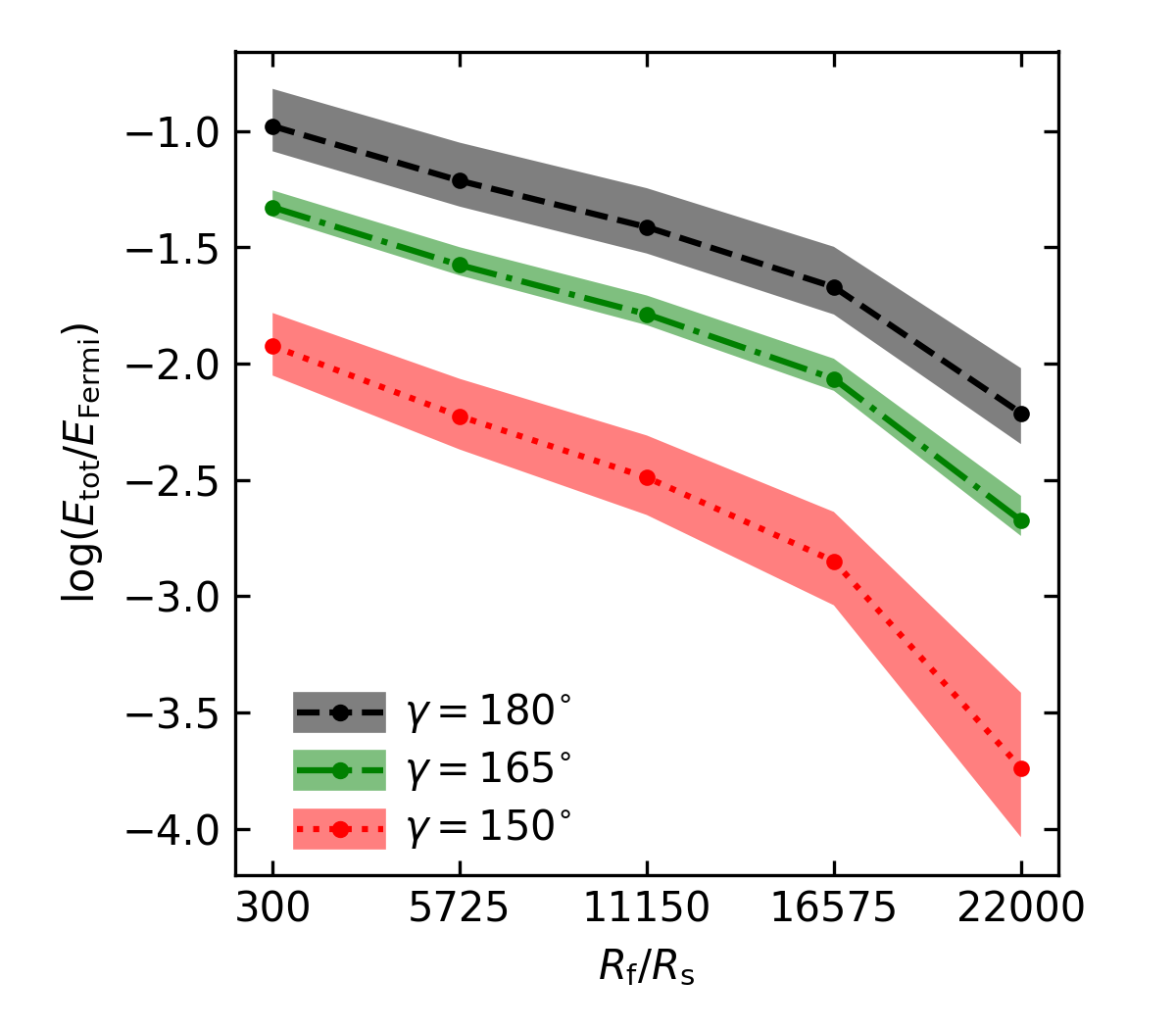}
	\vspace{-0.75cm}
    \caption{The total energy released during the activity period (lines and points). Shaded regions show variations among models with the same $\gamma$. }
    \label{fig:totalE}
\end{figure}

The actual accretion rate depends on mass transfer through the accretion disc, and is inevitably lower than the rate of mass transfer to the SMBH particle in the SPH simulation. The total energy released during the activity period, scaled to the energy required for {\em Fermi} bubble formation, is shown in fig. \ref{fig:totalE}. The total energy release required for the inflation of the {\em Fermi} bubbles is estimated to be $1.6\times 10^{58}$~erg \citep{kzbubles2012}, therefore accretion of $<10^5 \, \msun$ of gas is enough to produce them. We see that the total energy also decreases with increased $R_{\rm f}$. This happens because more of the gas leaves the system through the outer boundary as the feeding radius increases. Fig. \ref{fig:totalE} shows that models with collision angle of 180 degrees generate on average about $\sim 10\%$ of the energy required to form {\em Fermi} bubbles if gas is injected into the disc close to the black hole; most simulations with $\gamma \geq 165^\circ$ produce $> 1\%$ of the required energy.

We chose the minimal mass estimated by observations of the CNR for our initial system; this way we produce the lower limit for the accretion. The mass of the initial system could be larger by up to two orders of magnitude, i.e. $M_{\rm ring} = 10^6 \, \msun$. A cloud of this mass should have a radius of at least 5 pc even in the high-density environment of the Galactic Centre \citep{cloudSurvey}, therefore our setup is not completely valid for these cases. Nevertheless, as long as the cloud diameter is not much larger than the $2.5$~pc width of the ring, their collision involves the majority of the cloud's mass and a qualitatively similar evolution would be expected; in fact, a cloud less massive than the ring by a factor of a few might be enough to trigger an equivalent accretion episode (see Section \ref{colfreq}). As a result, the most massive possible system might produce 100 times more energy than in our simulations. 
An additional complication is that our accretion disc model is not well suited to model systems with luminosity reaching $L>L_{\rm{Edd}}$. Super-Eddington accretion causes some of the material to be shed before reaching the SMBH, therefore the difference in total released energy may not be as large as the increase in disc feeding rate. We return to this point in section \ref{feedback} of the Discussion. 

\section{Discussion }\label{Discussion}

Our models show that a collision between a CNR-like gas ring and a molecular cloud  with collision angle $\gamma \geq 105^{\circ}$ results in substantial gas transport to the centre of the system and, in more extreme cases, in the feeding of the central supermassive black hole. Small angle collisions produce the opposite results - increasing the extent and mass of the perturbed CNR. 

We begin by discussing the impact some chosen parameters - the gas cooling rate in section \ref{Betarez} and the ring-cloud mass ratio \ref{sec:relMass} - have on accretion. In section \ref{colfreq} we crudely estimate the expected frequency of collisions and the resulting AGN duty cycle. We discuss the implications of our findings for the Milky Way and other galaxies in sections \ref{evidence}-\ref{otherG}. The dense, possibly star forming rings found in some of the simulations and the challenge they pose to our initial assumptions is discussed in section  \ref{StellarRing}. We address the effects of feedback and super-Eddington accretion, neglected in our simulations, in section \ref{feedback}.

\subsection{Gas transport and the cooling of gas} \label{Betarez}
\begin{figure}
	\includegraphics[width=\columnwidth]{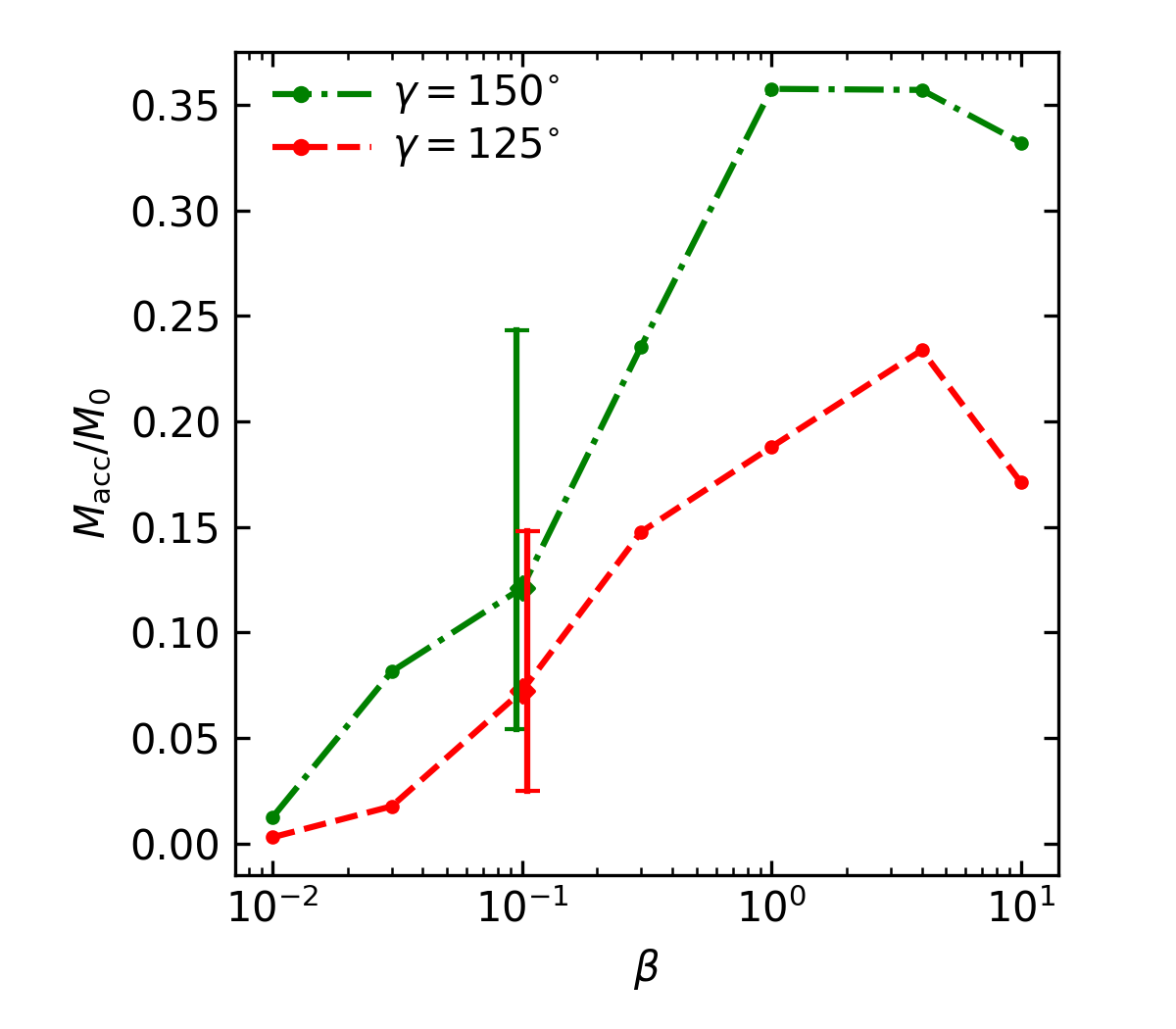}
	\vspace{-0.75cm}
    \caption{Relationship between the cooling time parameter $\beta$ and the total amount of accreted gas in units of $M_0 = 10^4 \msun$. The value of $\beta_{\rm c}$ used for the main set of models is marked with diamonds. The variations of accreted mass is shown as error bars.
    }
    \label{fig:Betacool}
\end{figure}

To parameterize the cooling of gas we use the beta-cooling prescription \citep{BETA2011} where the cooling time is $t_{\rm c} = \beta_{\rm c} t_{\rm dyn}$. This is a significant simplification of actual cooling processes, so in order to test the importance of cooling on our results, we ran several simulations varying $\beta_{\rm c}$. The results given in fig. \ref{fig:Betacool} show that longer cooling time for gas increases accretion up to a maximum factor $\sim3$ higher than fiducial models.

This happens because rapid cooling of gas results in reduced thickness of gas streams. As a result, collisions become rarer, making it more difficult to cancel out angular momentum, which results in reduced gas transport to the central regions. It is also important to note that the stochastic variation in total accreted mass (error bars in fig. \ref{fig:Betacool}) is as large as caused by a change of the value of $\beta_{\rm c}$ by a factor 3.

One piece of evidence for the activity period $\sim 6$~Myr ago is the ring of young stars around \sgra. They probably formed from a fragmenting gas disc and/or ring. It is known that fragmentation of rings in simulations requires the cooling parameter to be $<4.5 - 6$ \citep{SGRAFRAG}. Therefore, the value of $\beta_{\rm c}$ chosen for our main set of models is rather conservative, since more efficient cooling allows for less accretion. The short cooling times also compensate, to some extent, the lack of self-gravity in our simulations, by making the gas streams narrower than they might otherwise be.

\subsection{Accretion dependence on \MTc{cloud parameters}} \label{sec:relMass}
\begin{figure}
	\includegraphics[width=\columnwidth]{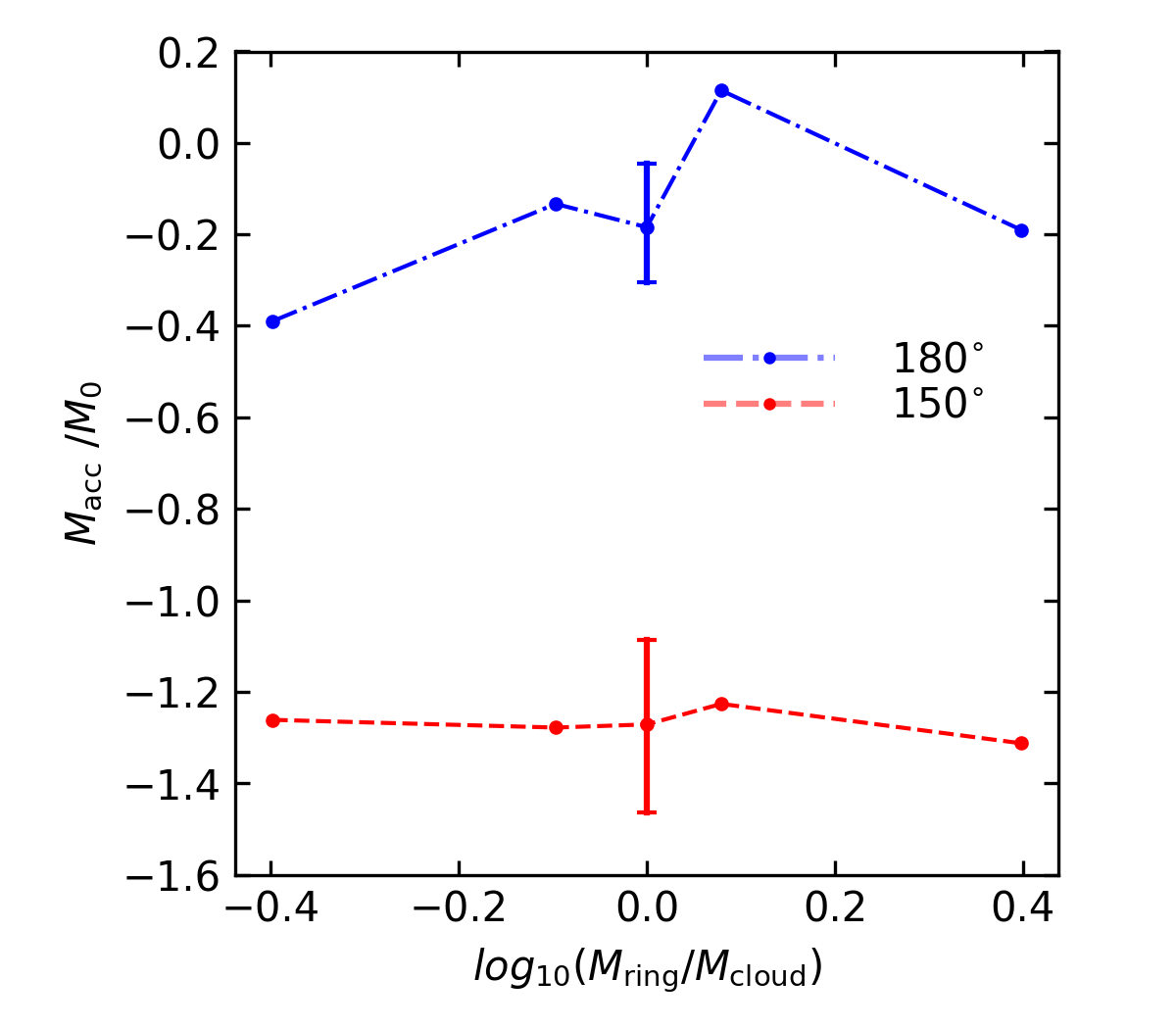}
	\vspace{-0.75cm}
    \caption{Total accreted mass relative to total initial mass of the system after $0.2\,$Myr for varied initial ring/cloud mass. The average and the extend of the main set of models is shown at the point $M_{\rm ring} = M_{\rm cloud}$. 
    }
    \label{fig:relMass}
\end{figure}

In our simulations, the masses of the ring and the cloud are the same, which is a contrived scenario. In order to determine how the this mass ratio affects the main results, we performed several simulations of collisions with different ring masses. Simulations were performed for the two most extreme collisions, $\gamma = 150\deg$ and $180\deg$, with mass ratios $M_{\rm cloud}/M_{\rm ring} = {0.4,\,0.8,\,1.2,\,2.5}$, while $M_{\rm cloud} = M_0 = 10^4\msun$ was kept the same as in the main simulations. The results of total mass transfer to the SMBH particles are shown in fig. \ref{fig:relMass}. 

\begin{figure}
	\includegraphics[width=\columnwidth]{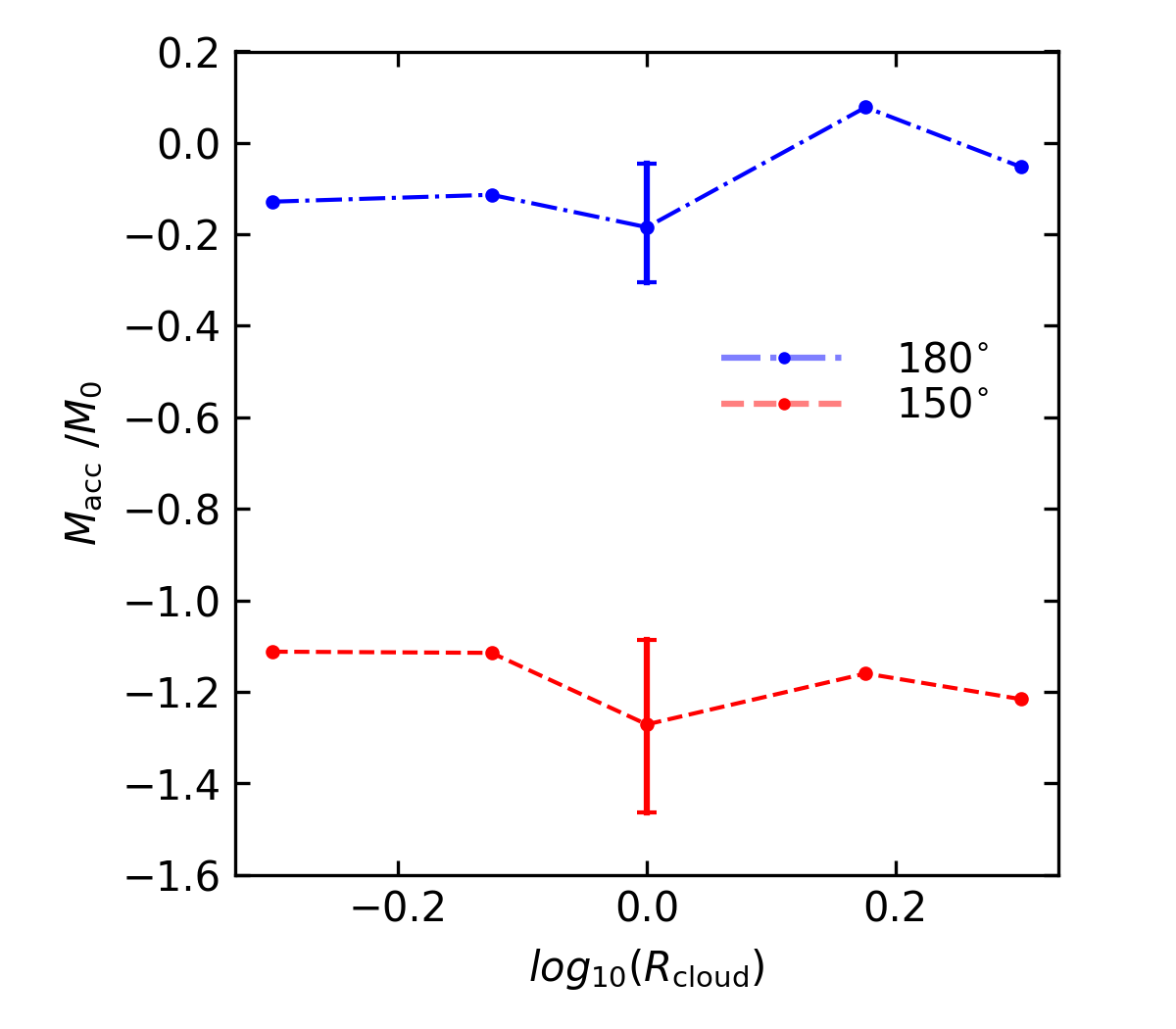}
	\vspace{-0.75cm}
    \caption{Total accreted mass relative to total initial mass of the system after $0.2\,$Myr for varied cloud radius. The average and the extend of the main set of models is shown at the point $R_{\rm ring} = 1$~pc. 
    }
    \label{fig:relR}
\end{figure}

\MTc{Similarly, we performed calculations varying the radius of the cloud. For these, we keep the masses of the ring and the cloud the same as in the main set of simulations. 
We ran calculations with cloud radius changed to $R_{\rm cloud}/{\rm pc} = {0.5,\,0.75,\,1.5,\,2}$. The results of total mass transfer to the SMBH particle are shown in fig. \ref{fig:relR}.
Note that simulations with varied cloud parameters encompassed only $0.2\,$Myr; we did not continue them further because the evolution of the system was very similar in all cases to the main simulations. The change in total accreted mass due to the different mass ratios or cloud radius is comparable to the variation caused by random initial distribution of particles (vertical error bars in the Figure), and thus can safely be neglected at the moment.}

\subsection{Frequency of collisions} \label{colfreq}

\begin{figure}
	\includegraphics[width=\columnwidth]{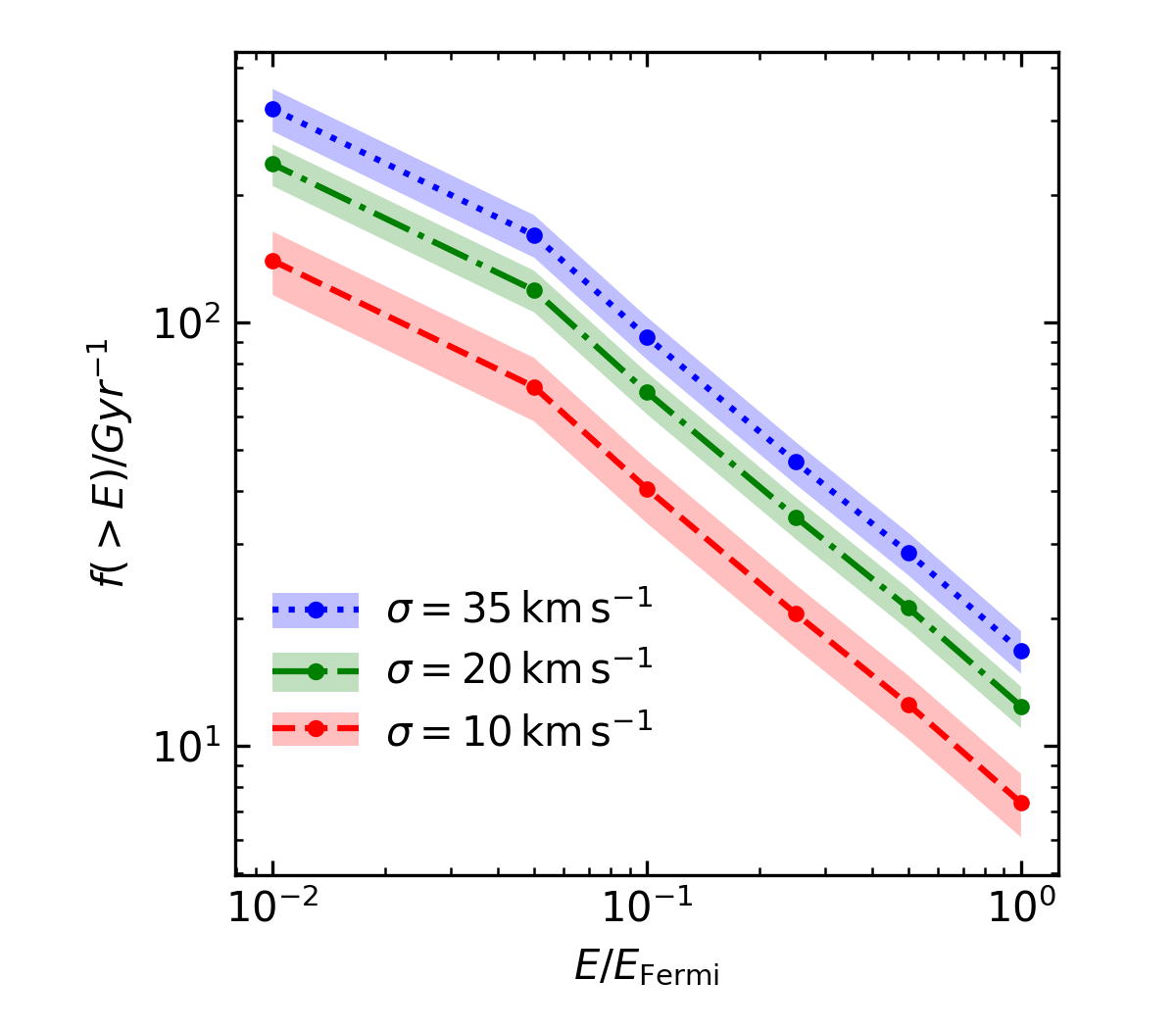}
	\vspace{-0.75cm}
    \caption{Estimated frequency of collisions that result in the release of at least $E/E_{\rm Fermi}$ energy, scaled to CMZ mass of $5\times10^7\,\msun$.}
    \label{fig:Freq}
\end{figure}

Our hydrodynamical simulations follow only a single event, but collisions should repeat stochastically in time as material flows into the CNR. This matter flow could build up the CNR resulting in a significant activity episode once an extreme collision occurs. We can make a rough estimate of the timescales of these processes by looking at how many clouds might have orbits that take them closer than $r_{\rm t} = 4\,$pc to the BH ($\sim r_{out}$ of the CNR) and how likely that cloud is to come in at an angle that results in significant accretion. Given the mass spectrum of the MCs \citep{massSpectrum} and the mass of the CMZ \citep[$M_{\rm CMZ} \approx 3 - 5\, \times 10^7\, \msun$][]{traces2012}, we generate a random uniform distribution of clouds in $x_1$ and $x_2$ orbits. The cloud velocity in the x-y plane is $v_{xy} = 165/\sqrt 2 \pm \sigma \,$km/s and in the z direction $v_{\rm z} = 0 \pm sigma $ km/s. We test three values for the velocity dispersion $\sigma = 10,\,20,\,35 \,$km~s$^{-1}$. We calculate the orbital components of each cloud and check if the cloud's periapsis is smaller than $r_{\rm t}$. The frequency is then given by the inverse of the orbital period for each cloud fulfilling said condition. 
A cloud of certain mass produces significant accretion only if the collision is steep enough; more massive clouds are more likely to produce significant accretion, because the range of viable angles is greater, but they are also less numerous. This is taken into account using eq. \ref{eq:fitted}, which relates the total accreted mass and collision angle, and the mass spectrum of the MCs which gives a number of clouds in a given mass range. Frequency is weighted by the `viable' range of angles of a given cloud.
After performing calculations with 100 random cloud distributions, we find that an activity period in which the SMBH would be fed by at least $\sim 4.5 \times 10^3\, \msun$ ($\sim 0.05E_{\rm Fermi}$) would occur every $3.1 - 7.2$~Myr, while a much larger collision that could result in {\em Fermi} bubble formation occurs every $60 - 140$~Myr (Fig. \ref{fig:Freq}). These results are given for a CMZ mass of $5\times10^7\,\msun$; the frequency of collisions scales linearly with CMZ mass. 

This is a very simplified calculation as we do not consider interactions between clouds and their orbital evolution. Also, equation (\ref{eq:fitted}) describes conditions where both CNR and MC are about the same mass. But many collisions occur that do not result in significant accretion and in effect only increase the mass of the CNR. Based on the same assumptions as above, we estimate the CNR growth rate to be $\dot{M}_{\rm CNR} \sim (4.3-6.0)\times10^{-2}\,\msun\,\rm{yr}^{-1}$, which is larger than the lower limit of the CNR mass growth $\dot{M}_{\rm min} \sim 2\times10^{-3}\,\msun\,\rm{yr}^{-1}$ \citep{CNRfeed2}.

\subsection{Evidence of past collisions in system morphology} \label{evidence}

Simulations of the formation of the CNR show that an infalling MC is a likely explanation of its origin \citep{CNRform2018}. But the inflow of gas into the centre does not stop after the formation of the CNR; instead, it continues with other MCs and gas streams falling in, resulting in collisions between the CNR-like gas ring at the centre and the observed infalling matter from larger scales \citep{CNRfeedl,CNRfeed2}. 

Our simulations show that these collisions do not necessarily result in significant nuclear activity  - looking at fig. \ref{fig:structMass} we can see that for $\gamma \leq 105^{\circ}$ almost all of the gas is still in the form of rings and its mass is larger than the initial mass of the CNR-like torus. Fig. \ref{fig:circOrb} shows that collisions with $90^{\circ} < \gamma < 120^{\circ}$ result in a system of similar extent to the initial ring. Collisions at a larger angle leave rings that are smaller than the initial one. Thus it seems plausible that the CNR-analogue that existed $>6$~Myr ago might have been more massive and larger than the present-day one, and was reduced in both mass and size following the extreme collision.

This process implies the existence of a cycle, where the ring system grows in mass, up until a collision feeds its mass to the black hole and/or star formation begins in the densest fragmenting regions. Additionally, star formation might not be confined to the central parsec. Dense filaments exist in the outer rings in our simulations. Provided that they are not disrupted by infalling matter, they could grow to be dense enough and create elongated stellar filaments. This could be contrasted with a single infalling cloud, which would result in star formation over a wider region \citep{StarFormationGC1}.

The CNR system, as it is seen today, provides some possible evidence of previous collisions. In the central cavity of the CNR there is a minispiral \citep{gas}. \cite{CNRform2018} hints that this central structure forms during the formation of the CNR. Our simulations suggest that similar structures could also form after collision with preexisting CNR. There are also cloudlets observed in the Central cavity which are possibly remnants of a smaller disruped gas cloud \citep{cloudlets}. These are short lived structures, which suggest that there is somewhat regural inflow of clouds.

\subsection{Similar events in other galaxies} \label{otherG}

Structures similar to those in the GC can be found in other galaxies, including AGN hosts\citep{traces2012, MW_AGN}. 
One of the observed features are the CMZ-like molecular gas rings\footnote{Often in literature the CMZ-scale features in other galaxies are also called Circumnuclear rings (CNRs)} on a scale of about 100 pc up to 2 kpc; galaxies that have molecular rings are more likely to be active \citep{NuclearinCNR}. A CNR-like feature is often unresolved, but there seem to be examples of both central molecular discs and filaments transporting gas closer to the centre \citep[e.g.,][]{circinus,NGC1097}. \MTc{More recently, observations with ALMA found a counter rotating $\sim1$~pc scale disc relative to the torus of the active galaxy NGC 1068 \citep{NGC1068}. This two-disc configuration suggests that the structure formed in more than one accretion event and is indirect evidence for the chaotic accretion scenario \citep{chaoticacc}.}
These observations suggest that CNR-like structures are closely connected to nuclear activity, perhaps through a cycle of growth and depletion via extreme collisions, as suggested by our simulations. 

At present, it is highly speculative to extrapolate collision frequencies for other galaxies as even the larger scale CMZ-like rings remain geometrically not very well, if at all, defined \citep{NuclearinCNR}. If we assume similarity between our galaxy and others that contain CMZ-scale structures, we can speculate that gas inflow to the centre from this CMZ should first produce a CNR-like parsec-scale structure with further collisions resulting in either a build up of gas in the CNR-like structure or a larger activity event on a timescale similar to the one calculated in Section \ref{colfreq} above, linearly scaled to the mass of the CMZ-like rings. As seen in fig. \ref{fig:structMass}, the CNR-like ring is significantly depleted after a collision resulting in significant accretion. But if we assume that new mass continually enters the system it would take $0.2\,$Myr to $2\,$Myr for the CNR to grow to $10^4\,\msun$. This is comparable to, but somewhat longer than, the timescale of the activity period (a few times $10^5$~yr). Therefore a CNR-like feature with a small mass could be an indicator of recent nuclear activity.

\subsection{Formation of stellar rings }\label{StellarRing}

The Galactic centre region contains a number of relatively young massive stars \citep[ZAMS mass between $30 - 100 \, \textrm{M}_{\odot}$; cf.][]{gcrev}. Most of the young stars located in the central $R\leq 0.5\,$pc formed in a single episode $6 \pm 2\,$Myr ago \citep{YoungDiscEvidence}. This colection of stars is modeled as two co-eval \citep{YoungWarpedDisc} discs rotating in clockwise and counter-clockwise directions, with a large angle between them \citep{gcrev}. \MTc{The existence of the counter-clockwise disc is in doubt as more recent studies of the young star cluster kinematics did not detect a significant feature \citep{YELDA}}.

A star formation episode could result from a fragmenting gas ring, possibly formed after a collision \citep{StarFormationGC1, StarFormationGC2}. 
\MTc{Possible star formation in two misaligned rings fallowing a collision between a cloud and the CNR was the focus of \cite{Alig2}. They've also found that such an event results in considerable accretion onto the SMBH.} In our simulations gas self-gravity is turned off, so we cannot track gas fragmentation and star formation directly. Even so, some rings separate at the outer edge of the central disc at $\sim1\,$pc (see fig. \ref{fig:Dens_hist} and left panel of fig \ref{fig:Slice_warp}), and most collisions form some rings that are narrower and denser than the initial one. The rings break apart as the particles travel closer to the centre, exchanging angular momentum with the faster moving particles.

As the mass of these rings grows, so does their density, reaching values above the tidal density. The Toomre Q parameter \citep{Qpar},
\begin{equation}
    Q = \left(\frac{GM}{r^3} + \frac{2\sigma^2}{r^2}\right) \frac{c_\textrm{s}}{\pi G \Sigma},
\end{equation}
is smaller than unity ($Q<1$) in some regions in simulations with $150\deg \leq \gamma \leq 165 \deg$. Here self-gravity is important and our initial assumptions are somewhat challenged, but the discs themselves remain comfortably below the tidal density, so our main results remain unaffected. These rings could in principle fragment and star formation could occur. While our rings form at about twice the distance ($\sim1$~pc) from the GC than the observed young stellar rings ($<0.5$~pc) it is interesting that a single collision may result in a possibly star forming  warped disc as well as a non-trivial amount of accretion giving more credence to the scenario where the birth of the young stars $6$~Myr ago is linked with the {\em Fermi} bubble formation.

\subsection{Effects of feedback} \label{feedback}

Our model neglects the effect of feedback from the accretion disc to the hydrodynamical system, but it is clear that some material escapes the accretion disc. Unfortunately, the particles used in the SPH simulation have comparable smoothing lengths to the diameter of the accretion disc, therefore we have no knowledge of the orientation of the disc w.r.t. our hydrodynamical system. 

However,even if feedback was directly striking the structures in the hydrodynamical simulation, it would have little effect. To see this, consider the following argument.

To see that a hypothetical outflow from \sgra \, would not have a significant effect on the CNR, consider the following argument. Taking the mass stored in the central discs (fig. \ref{fig:structMass}) we estimate their weights \citep{WEIGHTS}:
\begin{equation}
    W_{\rm disc} \sim g(R)M_{\rm disc}(<R),
\end{equation}
where $g(R)$ is the gravitational acceleration calculated at R from the potential of the system: 
\begin{equation}
g\left(R\right) = \frac{\textrm{G}M_{\rm{bh}}}{R^2} - \frac{2\sigma^2}{R} .
\label{gacc}
\end{equation}
The weight is $W_{\rm disc} \sim 5.4\times10^{35} M_4$~dyn at 1 pc from SMBH. We then estimate the outward force as:
\begin{equation}
    F_{\rm out} \sim \frac{H}{R}\frac{L_{\rm Edd}}{c} = 1.37 \times 10^{34} \frac{H}{R} \, {\rm dyn},
\end{equation}
where we take $H/R$ to be the largest geometrical ratio for the chosen disc, which in most cases is about $0.06-0.09$.

We find that in all cases, the weights of the discs are at least two orders of magnitude larger than the outward force. If we increase the initial mass of the system, the weight increases linearly, but the maximum luminosity only increases logarithmically above the Eddington limit. Therefore, feedback should become even less relevant for systems with higher initial mass. Energy-driven feedback should also have little impact on the resultant system, as most of it would escape through the less dense opening perpendicular to the disc \citep{Energydriven}. 

One more thing we have to consider is the limitations of our accretion disc model. We cannot accurately recreate activity with luminosities larger than the Eddington limit. This is a problem, because we cannot generate energies $>E_{\rm Fermi}$ without passing $L_{\rm Edd}$, even in the case of prolonged activity with large $R_{f}$ values. Finally, if the accretion rate through the disc becomes significantly super-Eddington, luminosity may be beamed \citep{BEAMING}, effectively producing more feedback in some directions. We intend to account for these effects, and include a self-consistent treatment of feedback, in future work.

\section{Conclusions}\label{Conclusions}

With a hydrodynamical model we show that a collision of the Circumnuclear Ring and a molecular cloud strongly perturbs the initial system; its subsequent evolution strongly depends on the initial collision angle between the orbit of the cloud and the plane of the ring. Collisions with angles $\gamma \leq105\deg$ result in the CNR increasing in size and mass, with minimal mass transfer to the central part of the system; those with $\gamma >105\deg$ result in an increased transport of gas toward the Galactic centre, where it forms a warped disc and feeds the central SMBH.

In the case of the most extreme collisions between a $10^5 \msun$ cloud and similar mass CNR, the mass transferred would be enough to power an AGN episode for more than 1 Myr. Such an episode would release enough energy to inflate the {\em Fermi} bubbles, while the self-gravitating outskirts of the accretion disc may form stars, creating the observed stellar rings. We estimate that a collision extreme enough to form the {\em Fermi} bubbles could happen once every $60 - 140$~Myr. Therefore, our model is a plausible explanation for the history of the accretion event that happened in our Galaxy $\sim 6$~Myr ago. Similar events may be ocurring in other galaxies, leaving footprints visible for several Myr afterwards.

\section*{Acknowledgements}

We thank Sergei Nayakshin for illuminating discussions and suggestions. This research was funded by the Research Council Lithuania grant no. MIP-17-78. Hydrodynamical simulations presented in this paper were carried out on the Vilnius University HPC cluster Fizika 1000.











\bsp	
\label{lastpage}

\begin{thebibliography}{}
\makeatletter
\relax
\def\mn@urlcharsother{\let\do\@makeother \do\$\do\&\do\#\do\^\do\_\do\%\do\~}
\def\mn@doi{\begingroup\mn@urlcharsother \@ifnextchar [ {\mn@doi@}
  {\mn@doi@[]}}
\def\mn@doi@[#1]#2{\def\@tempa{#1}\ifx\@tempa\@empty \href
  {http://dx.doi.org/#2} {doi:#2}\else \href {http://dx.doi.org/#2} {#1}\fi
  \endgroup}
\def\mn@eprint#1#2{\mn@eprint@#1:#2::\@nil}
\def\mn@eprint@arXiv#1{\href {http://arxiv.org/abs/#1} {{\tt arXiv:#1}}}
\def\mn@eprint@dblp#1{\href {http://dblp.uni-trier.de/rec/bibtex/#1.xml}
  {dblp:#1}}
\def\mn@eprint@#1:#2:#3:#4\@nil{\def\@tempa {#1}\def\@tempb {#2}\def\@tempc
  {#3}\ifx \@tempc \@empty \let \@tempc \@tempb \let \@tempb \@tempa \fi \ifx
  \@tempb \@empty \def\@tempb {arXiv}\fi \@ifundefined
  {mn@eprint@\@tempb}{\@tempb:\@tempc}{\expandafter \expandafter \csname
  mn@eprint@\@tempb\endcsname \expandafter{\@tempc}}}

\bibitem[\protect\citeauthoryear{{Alig}, {Burkert}, {Johansson}  \&
  {Schartmann}}{{Alig} et~al.}{2011}]{Alig1}
{Alig} C.,  {Burkert} A.,  {Johansson} P.~H.,   {Schartmann} M.,  2011, \mn@doi
  [\mnras] {10.1111/j.1365-2966.2010.17915.x}, \href
  {https://ui.adsabs.harvard.edu/abs/2011MNRAS.412..469A} {412, 469}

\bibitem[\protect\citeauthoryear{{Alig}, {Schartmann}, {Burkert}  \&
  {Dolag}}{{Alig} et~al.}{2013}]{Alig2}
{Alig} C.,  {Schartmann} M.,  {Burkert} A.,   {Dolag} K.,  2013, \mn@doi [\apj]
  {10.1088/0004-637X/771/2/119}, \href
  {https://ui.adsabs.harvard.edu/abs/2013ApJ...771..119A} {771, 119}

\bibitem[\protect\citeauthoryear{{Bartko} et~al.,}{{Bartko}
  et~al.}{2009}]{YoungWarpedDisc}
{Bartko} H.,  et~al., 2009, \mn@doi [\apj] {10.1088/0004-637X/697/2/1741},
  \href {https://ui.adsabs.harvard.edu/abs/2009ApJ...697.1741B} {697, 1741}

\bibitem[\protect\citeauthoryear{{Boehle} et~al.,}{{Boehle}
  et~al.}{2016}]{BHMASS}
{Boehle} A.,  et~al., 2016, \mn@doi [\apj] {10.3847/0004-637X/830/1/17}, \href
  {https://ui.adsabs.harvard.edu/abs/2016ApJ...830...17B} {830, 17}

\bibitem[\protect\citeauthoryear{Bonnell \& Rice}{Bonnell \&
  Rice}{2008}]{StarFormationGC1}
Bonnell I.~A.,  Rice W. K.~M.,  2008, \mn@doi [Science]
  {10.1126/science.1160653}, 321, 1060

\bibitem[\protect\citeauthoryear{{Christopher}, {Scoville}, {Stolovy}  \&
  {Yun}}{{Christopher} et~al.}{2005}]{maxmass}
{Christopher} M.~H.,  {Scoville} N.~Z.,  {Stolovy} S.~R.,   {Yun} M.~S.,  2005,
  \mn@doi [\apj] {10.1086/427911}, \href
  {http://adsabs.harvard.edu/abs/2005ApJ...622..346C} {622, 346}

\bibitem[\protect\citeauthoryear{{Dehnen} \& {Aly}}{{Dehnen} \&
  {Aly}}{2012}]{kernel}
{Dehnen} W.,  {Aly} H.,  2012, \mn@doi [\mnras]
  {10.1111/j.1365-2966.2012.21439.x}, \href
  {http://adsabs.harvard.edu/abs/2012MNRAS.425.1068D} {425, 1068}

\bibitem[\protect\citeauthoryear{{Dubinski}, {Narayan}  \&
  {Phillips}}{{Dubinski} et~al.}{1995}]{Turbulencija}
{Dubinski} J.,  {Narayan} R.,   {Phillips} T.~G.,  1995, \mn@doi [\apj]
  {10.1086/175954}, \href
  {https://ui.adsabs.harvard.edu/\#abs/1995ApJ...448..226D} {448, 226}

\bibitem[\protect\citeauthoryear{{Ferri{\`e}re}}{{Ferri{\`e}re}}{2012}]{gas}
{Ferri{\`e}re} K.,  2012, \mn@doi [\aap] {10.1051/0004-6361/201117181}, \href
  {http://adsabs.harvard.edu/abs/2012A%26A...540A..50F} {540, A50}

\bibitem[\protect\citeauthoryear{Frank, King  \& Raine}{Frank
  et~al.}{2002}]{frank_king_raine_2002}
Frank J.,  King A.,   Raine D.,  2002, Accretion Power in Astrophysics, 3 edn.
Cambridge University Press, \mn@doi{10.1017/CBO9781139164245}

\bibitem[\protect\citeauthoryear{{Fritz} et~al.,}{{Fritz}
  et~al.}{2016}]{NuclearStarPot}
{Fritz} T.~K.,  et~al., 2016, \mn@doi [\apj] {10.3847/0004-637X/821/1/44},
  \href {https://ui.adsabs.harvard.edu/abs/2016ApJ...821...44F} {821, 44}

\bibitem[\protect\citeauthoryear{{Genzel}, {Eisenhauer}  \&
  {Gillessen}}{{Genzel} et~al.}{2010}]{gcrev}
{Genzel} R.,  {Eisenhauer} F.,   {Gillessen} S.,  2010, \mn@doi [Reviews of
  Modern Physics] {10.1103/RevModPhys.82.3121}, \href
  {http://adsabs.harvard.edu/abs/2010RvMP...82.3121G} {82, 3121}

\bibitem[\protect\citeauthoryear{{Goicoechea}, {Pety}, {Chapillon},
  {Cernicharo}, {Gerin}, {Herrera}, {Requena-Torres}  \&
  {Santa-Maria}}{{Goicoechea} et~al.}{2018}]{cloudlets}
{Goicoechea} J.~R.,  {Pety} J.,  {Chapillon} E.,  {Cernicharo} J.,  {Gerin} M.,
   {Herrera} C.,  {Requena-Torres} M.~A.,   {Santa-Maria} M.~G.,  2018, \mn@doi
  [\aap] {10.1051/0004-6361/201833558}, \href
  {https://ui.adsabs.harvard.edu/\#abs/2018A&A...618A..35G} {618, A35}

\bibitem[\protect\citeauthoryear{{Heywood} et~al.,}{{Heywood}
  et~al.}{2019}]{430RADIOBUBBLES}
{Heywood} I.,  et~al., 2019, \mn@doi [\nat] {10.1038/s41586-019-1532-5}, \href
  {https://ui.adsabs.harvard.edu/abs/2019Natur.573..235H} {573, 235}

\bibitem[\protect\citeauthoryear{{Hobbs} \& {Nayakshin}}{{Hobbs} \&
  {Nayakshin}}{2009}]{StarFormationGC2}
{Hobbs} A.,  {Nayakshin} S.,  2009, \mn@doi [\mnras]
  {10.1111/j.1365-2966.2008.14359.x}, \href
  {https://ui.adsabs.harvard.edu/abs/2009MNRAS.394..191H} {394, 191}

\bibitem[\protect\citeauthoryear{{Hobbs}, {Nayakshin}, {Power}  \&
  {King}}{{Hobbs} et~al.}{2011}]{ACC_RING_angularmomentum}
{Hobbs} A.,  {Nayakshin} S.,  {Power} C.,   {King} A.,  2011, \mn@doi [\mnras]
  {10.1111/j.1365-2966.2011.18333.x}, \href
  {https://ui.adsabs.harvard.edu/abs/2011MNRAS.413.2633H} {413, 2633}

\bibitem[\protect\citeauthoryear{{Hsieh}, {Koch}, {Ho}, {Kim}, {Tang}, {Wang},
  {Yen}  \& {Hwang}}{{Hsieh} et~al.}{2017}]{CNRfeed2}
{Hsieh} P.-Y.,  {Koch} P.~M.,  {Ho} P. T.~P.,  {Kim} W.-T.,  {Tang} Y.-W.,
  {Wang} H.-H.,  {Yen} H.-W.,   {Hwang} C.-Y.,  2017, \mn@doi [\apj]
  {10.3847/1538-4357/aa8329}, \href
  {https://ui.adsabs.harvard.edu/abs/2017ApJ...847....3H} {847, 3}

\bibitem[\protect\citeauthoryear{{Impellizzeri} et~al.,}{{Impellizzeri}
  et~al.}{2019}]{NGC1068}
{Impellizzeri} C.~M.~V.,  et~al., 2019, \mn@doi [\apjl]
  {10.3847/2041-8213/ab3c64}, \href
  {https://ui.adsabs.harvard.edu/abs/2019ApJ...884L..28I} {884, L28}

\bibitem[\protect\citeauthoryear{{Issaoun} et~al.,}{{Issaoun}
  et~al.}{2019}]{SizeShapeSgrA}
{Issaoun} S.,  et~al., 2019, \mn@doi [\apj] {10.3847/1538-4357/aaf732}, \href
  {https://ui.adsabs.harvard.edu/abs/2019ApJ...871...30I} {871, 30}

\bibitem[\protect\citeauthoryear{{Izumi}, {Wada}, {Fukushige}, {Hamamura}  \&
  {Kohno}}{{Izumi} et~al.}{2018}]{circinus}
{Izumi} T.,  {Wada} K.,  {Fukushige} R.,  {Hamamura} S.,   {Kohno} K.,  2018,
  \mn@doi [\apj] {10.3847/1538-4357/aae20b}, \href
  {https://ui.adsabs.harvard.edu/abs/2018ApJ...867...48I} {867, 48}

\bibitem[\protect\citeauthoryear{{Kauffmann}, {Pillai}, {Zhang}, {Menten},
  {Goldsmith}, {Lu}  \& {Guzm{\'a}n}}{{Kauffmann} et~al.}{2017}]{cloudSurvey}
{Kauffmann} J.,  {Pillai} T.,  {Zhang} Q.,  {Menten} K.~M.,  {Goldsmith} P.~F.,
   {Lu} X.,   {Guzm{\'a}n} A.~E.,  2017, \mn@doi [\aap]
  {10.1051/0004-6361/201628088}, \href
  {https://ui.adsabs.harvard.edu/\#abs/2017A&A...603A..89K} {603, A89}

\bibitem[\protect\citeauthoryear{{King}}{{King}}{2009}]{BEAMING}
{King} A.~R.,  2009, \mn@doi [Monthly Notices of the Royal Astronomical
  Society] {10.1111/j.1745-3933.2008.00594.x}, \href
  {https://ui.adsabs.harvard.edu/abs/2009MNRAS.393L..41K} {393, L41}

\bibitem[\protect\citeauthoryear{{King} \& {Pringle}}{{King} \&
  {Pringle}}{2007}]{chaoticacc}
{King} A.~R.,  {Pringle} J.~E.,  2007, \mn@doi [\mnras]
  {10.1111/j.1745-3933.2007.00296.x}, \href
  {https://ui.adsabs.harvard.edu/abs/2007MNRAS.377L..25K} {377, L25}

\bibitem[\protect\citeauthoryear{Liu, Hsieh, Ho, Su, Wright, Sun  \&
  Chol~Minh}{Liu et~al.}{2012}]{CNRfeedl}
Liu H.,  Hsieh P.-Y.,  Ho P.,  Su Y.-N.,  Wright M.,  Sun A.-L.,   Chol~Minh
  Y.,  2012, \mn@doi [Astrophysical Journal] {10.1088/0004-637X/756/2/195}, 756

\bibitem[\protect\citeauthoryear{{Lucas}, {Bonnell}, {Davies}  \&
  {Rice}}{{Lucas} et~al.}{2013}]{LUCASMCINFALL}
{Lucas} W.~E.,  {Bonnell} I.~A.,  {Davies} M.~B.,   {Rice} W.~K.~M.,  2013,
  \mn@doi [\mnras] {10.1093/mnras/stt727}, \href
  {https://ui.adsabs.harvard.edu/abs/2013MNRAS.433..353L} {433, 353}

\bibitem[\protect\citeauthoryear{Mapelli \& Trani}{Mapelli \&
  Trani}{2015}]{CNRform2015}
Mapelli M.,  Trani A.,  2015, \mn@doi [Astronomy & Astrophysics]
  {10.1051/0004-6361/201527195}, 585

\bibitem[\protect\citeauthoryear{{Meru} \& {Bate}}{{Meru} \&
  {Bate}}{2011}]{BETA2011}
{Meru} F.,  {Bate} M.~R.,  2011, \mn@doi [\mnras]
  {10.1111/j.1745-3933.2010.00978.x}, \href
  {http://adsabs.harvard.edu/abs/2011MNRAS.411L...1M} {411, L1}

\bibitem[\protect\citeauthoryear{{Nayakshin}, {Cuadra}  \&
  {Springel}}{{Nayakshin} et~al.}{2007}]{SGRAFRAG}
{Nayakshin} S.,  {Cuadra} J.,   {Springel} V.,  2007, \mn@doi [Monthly Notices
  of the Royal Astronomical Society] {10.1111/j.1365-2966.2007.11938.x}, \href
  {https://ui.adsabs.harvard.edu/abs/2007MNRAS.379...21N} {379, 21}

\bibitem[\protect\citeauthoryear{{Nealon}, {Nixon}, {Price}  \&
  {King}}{{Nealon} et~al.}{2016}]{Warped_disc}
{Nealon} R.,  {Nixon} C.,  {Price} D.~J.,   {King} A.,  2016, \mn@doi [\mnras]
  {10.1093/mnrasl/slv149}, \href
  {https://ui.adsabs.harvard.edu/abs/2016MNRAS.455L..62N} {455, L62}

\bibitem[\protect\citeauthoryear{{Oka}, {Nagai}, {Kamegai}  \& {Tanaka}}{{Oka}
  et~al.}{2011}]{CNRVirialmass}
{Oka} T.,  {Nagai} M.,  {Kamegai} K.,   {Tanaka} K.,  2011, \mn@doi [\apj]
  {10.1088/0004-637X/732/2/120}, \href
  {https://ui.adsabs.harvard.edu/abs/2011ApJ...732..120O} {732, 120}

\bibitem[\protect\citeauthoryear{P.~Aguero, J.~Díaz  \& Dottori}{P.~Aguero
  et~al.}{2016}]{NuclearinCNR}
P.~Aguero M.,  J.~Díaz R.,   Dottori H.,  2016, \mn@doi [International Journal
  of Astronomy and Astrophysics] {10.4236/ijaa.2016.63018}, 06

\bibitem[\protect\citeauthoryear{{Paumard} et~al.,}{{Paumard}
  et~al.}{2006}]{YoungDiscEvidence}
{Paumard} T.,  et~al., 2006, \mn@doi [\apj] {10.1086/503273}, \href
  {https://ui.adsabs.harvard.edu/abs/2006ApJ...643.1011P} {643, 1011}

\bibitem[\protect\citeauthoryear{{Ponti}, {Morris}, {Terrier}  \&
  {Goldwurm}}{{Ponti} et~al.}{2013}]{traces2012}
{Ponti} G.,  {Morris} M.~R.,  {Terrier} R.,   {Goldwurm} A.,  2013, in {Torres}
  D.~F.,  {Reimer} O.,  eds,  Astrophysics and Space Science Proceedings Vol.
  34, Cosmic Rays in Star-Forming Environments. p.~331 (\mn@eprint {arXiv}
  {1210.3034}), \mn@doi{10.1007/978-3-642-35410-6_26}

\bibitem[\protect\citeauthoryear{{Ponti} et~al.,}{{Ponti}
  et~al.}{2019}]{XRAYCHIMNEY}
{Ponti} G.,  et~al., 2019, \mn@doi [\nat] {10.1038/s41586-019-1009-6}, \href
  {https://ui.adsabs.harvard.edu/abs/2019Natur.567..347P} {567, 347}

\bibitem[\protect\citeauthoryear{Prieto, Fernandez-Ontiveros, Bruzual, Burkert,
  Schartmann  \& Charlot}{Prieto et~al.}{2019}]{NGC1097}
Prieto M.,  Fernandez-Ontiveros J.,  Bruzual G.,  Burkert A.,  Schartmann M.,
  Charlot S.,  2019, \mn@doi [Monthly Notices of the Royal Astronomical
  Society] {10.1093/mnras/stz579}, 485, 3264

\bibitem[\protect\citeauthoryear{{Pringle}}{{Pringle}}{1981}]{Pringle1981}
{Pringle} J.~E.,  1981, \mn@doi [\araa] {10.1146/annurev.aa.19.090181.001033},
  \href {http://adsabs.harvard.edu/abs/1981ARA%26A..19..137P} {19, 137}

\bibitem[\protect\citeauthoryear{{Read} \& {Hayfield}}{{Read} \&
  {Hayfield}}{2012}]{SPHS2012}
{Read} J.~I.,  {Hayfield} T.,  2012, \mn@doi [\mnras]
  {10.1111/j.1365-2966.2012.20819.x}, \href
  {http://adsabs.harvard.edu/abs/2012MNRAS.422.3037R} {422, 3037}

\bibitem[\protect\citeauthoryear{{Shakura} \& {Sunyaev}}{{Shakura} \&
  {Sunyaev}}{1973}]{ShakuraSunyaev}
{Shakura} N.~I.,  {Sunyaev} R.~A.,  1973, \aap, \href
  {http://adsabs.harvard.edu/abs/1973A%26A....24..337S} {24, 337}

\bibitem[\protect\citeauthoryear{{Springel}}{{Springel}}{2005}]{GADGET2005}
{Springel} V.,  2005, \mn@doi [\mnras] {10.1111/j.1365-2966.2005.09655.x},
  \href {http://adsabs.harvard.edu/abs/2005MNRAS.364.1105S} {364, 1105}

\bibitem[\protect\citeauthoryear{{Storchi-Bergmann}}{{Storchi-Bergmann}}{2014}]{MW_AGN}
{Storchi-Bergmann} T.,  2014, in {Sjouwerman} L.~O.,  {Lang} C.~C.,   {Ott} J.,
   eds,  IAU Symposium Vol. 303, The Galactic Center: Feeding and Feedback in a
  Normal Galactic Nucleus. pp 354--363 (\mn@eprint {arXiv} {1401.0032}),
  \mn@doi{10.1017/S174392131400091X}

\bibitem[\protect\citeauthoryear{{Su}, {Slatyer}  \& {Finkbeiner}}{{Su}
  et~al.}{2010}]{fermior}
{Su} M.,  {Slatyer} T.~R.,   {Finkbeiner} D.~P.,  2010, \mn@doi [\apj]
  {10.1088/0004-637X/724/2/1044}, \href
  {http://adsabs.harvard.edu/abs/2010ApJ...724.1044S} {724, 1044}

\bibitem[\protect\citeauthoryear{{Toomre}}{{Toomre}}{1964}]{Qpar}
{Toomre} A.,  1964, \mn@doi [\apj] {10.1086/147861}, \href
  {http://adsabs.harvard.edu/abs/1964ApJ...139.1217T} {139, 1217}

\bibitem[\protect\citeauthoryear{{Trani}, {Mapelli}  \& {Ballone}}{{Trani}
  et~al.}{2018}]{CNRform2018}
{Trani} A.~A.,  {Mapelli} M.,   {Ballone} A.,  2018, \mn@doi [\apj]
  {10.3847/1538-4357/aad414}, \href
  {https://ui.adsabs.harvard.edu/\#abs/2018ApJ...864...17T} {864, 17}

\bibitem[\protect\citeauthoryear{Williams \& Mckee}{Williams \&
  Mckee}{1997}]{massSpectrum}
Williams J.,  Mckee C.,  1997, \mn@doi [Astrophysical Journal v.476]
  {10.1086/303588}, 476, 166

\bibitem[\protect\citeauthoryear{{Yelda}, {Ghez}, {Lu}, {Do}, {Meyer}, {Morris}
   \& {Matthews}}{{Yelda} et~al.}{2014}]{YELDA}
{Yelda} S.,  {Ghez} A.~M.,  {Lu} J.~R.,  {Do} T.,  {Meyer} L.,  {Morris} M.~R.,
    {Matthews} K.,  2014, \mn@doi [\apj] {10.1088/0004-637X/783/2/131}, \href
  {https://ui.adsabs.harvard.edu/abs/2014ApJ...783..131Y} {783, 131}

\bibitem[\protect\citeauthoryear{{Zubovas} \& {Nayakshin}}{{Zubovas} \&
  {Nayakshin}}{2012}]{kzbubles2012}
{Zubovas} K.,  {Nayakshin} S.,  2012, \mn@doi [\mnras]
  {10.1111/j.1365-2966.2012.21250.x}, \href
  {http://adsabs.harvard.edu/abs/2012MNRAS.424..666Z} {424, 666}

\bibitem[\protect\citeauthoryear{{Zubovas} \& {Nayakshin}}{{Zubovas} \&
  {Nayakshin}}{2014}]{Energydriven}
{Zubovas} K.,  {Nayakshin} S.,  2014, \mn@doi [Monthly Notices of the Royal
  Astronomical Society] {10.1093/mnras/stu431}, \href
  {https://ui.adsabs.harvard.edu/abs/2014MNRAS.440.2625Z} {440, 2625}

\bibitem[\protect\citeauthoryear{{Zubovas}, {King}  \& {Nayakshin}}{{Zubovas}
  et~al.}{2011}]{WEIGHTS}
{Zubovas} K.,  {King} A.~R.,   {Nayakshin} S.,  2011, \mn@doi [Monthly Notices
  of the Royal Astronomical Society] {10.1111/j.1745-3933.2011.01070.x}, \href
  {https://ui.adsabs.harvard.edu/abs/2011MNRAS.415L..21Z} {415, L21}

\makeatother
\end{thebibliography}
\end{document}